\documentclass[twocolumn]{aastex63}
\let\tablenum\relax

\usepackage{amsmath} 
\usepackage{siunitx}
\usepackage{threeparttable}
\usepackage{CJK}

\received{XXX}
\revised{YYY}
\accepted{ZZZ}

\submitjournal{ApJ}

\shorttitle{C3D broad-line AGNs}
\shortauthors{Lin et al.}

\graphicspath{{./}{figures/}}

\newcommand{\ha}{H$\alpha$}
\newcommand{\hb}{H$\beta$}

\newcommand{\MBH}{$M_{\rm BH}$}

\newcommand{\betaopt}{$\beta_{\rm opt}$}

\newcommand{\hbagn}{C3D-z7AGN-1}

\begin{document}
\begin{CJK*}{UTF8}{gbsn}

\title{Bridging Quasars and Little Red Dots:\\ Insights into Broad-Line AGNs at $z$=5--8 from the First JWST COSMOS-3D Dataset}

\suppressAffiliations 
 

\author[0000-0001-6052-4234]{Xiaojing Lin}
\affiliation{Department of Astronomy, Tsinghua University, Beijing 100084, China}
\email{xiaojinglin.astro@gmail.com}
\affil{Steward Observatory, University of Arizona, 933 N Cherry Ave, Tucson, AZ 85721, USA}

\author[0000-0003-3310-0131]{Xiaohui Fan}
\affiliation{Steward Observatory, University of Arizona, 933 N Cherry Ave, Tucson, AZ 85721, USA}

\author[0000-0002-7633-431X]{Feige Wang}
\affiliation{Department of Astronomy, University of Michigan, 1085 S. University Ave., Ann Arbor, MI 48109, USA}

\author[0000-0002-4622-6617]{Fengwu Sun}
\affiliation{Center for Astrophysics $|$ Harvard \& Smithsonian, 60 Garden St., Cambridge, MA 02138, USA}


\author[0000-0002-6184-9097]{Jaclyn~B.~Champagne}
\affiliation{Steward Observatory, University of Arizona, 933 N Cherry Ave, Tucson, AZ 85721, USA}

\author[0000-0003-1344-9475]{Eiichi Egami}
\affiliation{Steward Observatory, University of Arizona, 933 N Cherry Ave, Tucson, AZ 85721, USA}

\author[0000-0001-6874-1321]{Koki Kakiichi}
\affiliation{Cosmic Dawn Center (DAWN), Denmark}
\affiliation{Niels Bohr Institute, University of Copenhagen, Jagtvej 128, DK-2200 Copenhagen N, Denmark}

\author[0000-0002-6221-1829]{Jianwei Lyu}
\affiliation{Steward Observatory, University of Arizona, 933 N Cherry Ave, Tucson, AZ 85721, USA}

\author[0000-0003-0747-1780]{Wei Leong Tee
}
\affiliation{Steward Observatory, University of Arizona, 933 N Cherry Ave, Tucson, AZ 85721, USA}

\author[0000-0001-5287-4242]{Jinyi Yang}
\affiliation{Department of Astronomy, University of Michigan, 1085 S. University Ave., Ann Arbor, MI 48109, USA}


\author[0000-0002-1620-0897]{Fuyan Bian}
\affiliation{European Southern Observatory, Alonso de C\'ordova 3107, Casilla 19001, Vitacura, Santiago 19, Chile}

\author[0000-0001-8582-7012]{Sarah E.~I.~Bosman}
\affiliation{Institute for Theoretical Physics, Heidelberg University, Philosophenweg 12, D-69120, Heidelberg, Germany}
\affiliation{Max-Planck-Institut f\"{u}r Astronomie, K\"{o}nigstuhl 17, 69117 Heidelberg, Germany}

\author[0000-0001-8467-6478]{Zheng Cai}
\affiliation{Department of Astronomy, Tsinghua University, Beijing 100084, China}

\author[0000-0002-0930-6466]{Caitlin M. Casey}
\affiliation{Department of Physics, University of California, Santa Barbara, Santa Barbara, CA 93106, USA}
\affiliation{The University of Texas at Austin, 2515 Speedway Blvd Stop C1400, Austin, TX 78712, USA}
\affiliation{Cosmic Dawn Center (DAWN), Denmark}

\author[0000-0002-2662-8803]{Roberto Decarli} \affiliation{INAF -- Osservatorio di Astrofisica e Scienza dello Spazio di Bologna, via Gobetti 93/3, I-40129, Bologna, Italy}

\author[0000-0002-9382-9832]{Andreas L. Faisst}
\affiliation{Caltech/IPAC, 1200 E. California Blvd. Pasadena, CA 91125, USA}

\author[0000-0001-8519-1130]{Steven L. Finkelstein}
\affiliation{Department of Astronomy, The University of Texas at Austin, Austin, TX 78712, USA}

\author[0000-0001-7201-5066]{Seiji Fujimoto}
\affiliation{
David A. Dunlap Department of Astronomy and Astrophysics, University of Toronto, 50 St. George Street, Toronto, Ontario, M5S 3H4, Canada
}
\affiliation{
Dunlap Institute for Astronomy and Astrophysics, 50 St. George Street, Toronto, Ontario, M5S 3H4, Canada
}

\author[0000-0003-0129-2079]{Santosh Harish}
\affiliation{Laboratory for Multiwavelength Astrophysics, School of Physics and Astronomy, Rochester Institute of Technology, 84 Lomb Memorial Drive, Rochester, NY 14623, USA}

\author[0000-0002-7303-4397]{Olivier Ilbert}
\affiliation{Aix Marseille Univ, CNRS, LAM, Laboratoire d'Astrophysique de Marseille, Marseille, France  \label{LAM} }

\author[0000-0002-7779-8677]{Akio K. Inoue}
\affiliation{Department of Physics, School of Advanced Science and Engineering, Faculty of Science and Engineering, Waseda University, 3-4-1, Okubo, Shinjuku, Tokyo 169-8555, Japan}
\affiliation{Waseda Research Institute for Science and Engineering, Faculty of Science and Engineering, Waseda University, 3-4-1, Okubo, Shinjuku, Tokyo 169-8555, Japan}

 \author[0000-0002-5768-738X]{Xiangyu Jin}
\affiliation{Steward Observatory, University of Arizona, 933 N Cherry Ave, Tucson, AZ 85721, USA}

\author[0000-0001-9187-3605]{Jeyhan S. Kartaltepe}
\affiliation{Laboratory for Multiwavelength Astrophysics, School of Physics and Astronomy, Rochester Institute of Technology, 84 Lomb Memorial Drive, Rochester, NY 14623, USA}

\author[0000-0002-8360-3880]{Dale D. Kocevski}
\affiliation{Department of Physics and Astronomy, Colby College, Waterville, ME 04901, USA}

\author[0000-0001-6251-649X]{Mingyu Li}
\affiliation{Department of Astronomy, Tsinghua University, Beijing 100084, China}

\author[0000-0003-3762-7344]{Weizhe Liu \begin{CJK}{UTF8}{gbsn}(刘伟哲)\end{CJK}}
\affiliation{Steward Observatory, University of Arizona, 933 N Cherry Ave, Tucson, AZ 85721, USA}

\author[0000-0003-4247-0169]{Yichen Liu}
\affiliation{Steward Observatory, University of Arizona, 933 N Cherry Ave, Tucson, AZ 85721, USA}

\author[0000-0002-4544-8242]{Jan-Torge Schindler}
\affiliation{Hamburger Sternwarte, Universität Hamburg, Gojenbergsweg 112, D-21029 Hamburg, Germany}

\author[0000-0002-7087-0701]{Marko Shuntov}
\affiliation{Cosmic Dawn Center (DAWN), Denmark}
\affiliation{Niels Bohr Institute, University of Copenhagen, Jagtvej 128, DK-2200 Copenhagen N, Denmark}
\affiliation{Department of Astronomy, University of Geneva, Chemin Pegasi 51, 1290 Versoix, Switzerland}

\author[0009-0003-4742-7060]{Takumi S. Tanaka}
\affiliation{Department of Astronomy, Graduate School of Science, The University of Tokyo, 7-3-1 Hongo, Bunkyo-ku, Tokyo, 113-0033, Japan}
\affiliation{Kavli Institute for the Physics and Mathematics of the Universe (WPI), The University of Tokyo Institutes for Advanced Study, The University of Tokyo, Kashiwa, Chiba 277-8583, Japan}
\affiliation{Center for Data-Driven Discovery, Kavli IPMU (WPI), UTIAS, The University of Tokyo, Kashiwa, Chiba 277-8583, Japan}

\author[0000-0001-9191-9837]{Marianne Vestergaard}
\affiliation{DARK, The Niels Bohr Institute, Jagtvej 155, 2200 Copenhagen N, Denmark}
\affiliation{Steward Observatory, University of Arizona, 933 N Cherry Ave, Tucson, AZ 85721, USA}

\author[0000-0003-0111-8249]{Yunjing Wu}
\affiliation{Department of Astronomy, Tsinghua University, Beijing 100084, China}

\author[0000-0002-4321-3538]{Haowen Zhang}
\affiliation{Steward Observatory, University of Arizona, 933 N Cherry Ave, Tucson, AZ 85721, USA}

\author[0000-0002-2420-5022]{Zijian Zhang}
\affiliation{Kavli Institute for Astronomy and Astrophysics, Peking University, Beijing 100871, China}
\affiliation{Department of Astronomy, School of Physics, Peking University, Beijing 100871, China}

\collaboration{99}{}
\collaboration{0}{(Affiliations can be found after the references)}

\correspondingauthor{Xiaojing Lin}

\begin{abstract}
We report the discovery of 13 broad-line AGNs at $z = 5 - 8$ from the first 10\% of JWST Cycle 3 Treasury Program COSMOS-3D. These AGNs are identified by their broad H$\alpha$ or H$\beta$ emission lines through NIRCam grism slitless spectroscopy. One object at $z = 7.646$ with broad H$\beta$ emission has an F444W magnitude of 23.6 mag, making it one of the brightest $z > 7.5$ broad-line AGNs yet known. Among the 13 AGNs, ten objects exhibit reddened optical continua with slopes $\beta_{\rm opt} > 0$.  The remaining three resemble UV-luminous quasars at similar redshift but with $\beta_{\rm opt}$ less blue than those of typical unobscured quasars. We also obtain MIRI photometry (7.7-18\,\micron) for two AGNs and place strong constraints on their rest-frame near-IR SED. We find no significant variability in the rest-frame UV by comparing the COSMOS-3D and COSMOS-Web F115W images taken apart by 60 days in the rest-frame. We compute the H$\alpha$ luminosity function (LF) at $z \approx 5-6$ and find potential redshift evolution compared to $z \approx 4-5$. We also derive the H$\beta$ LF at $z \sim 8$ by combining our sample with those from the literature. The broad H$\beta$  emitters in this work suggest a number density two orders of magnitude higher than that predicted by the quasar LF based on rest-frame UV-selected samples. As a preview, our work showcases the ability of the COSMOS-3D grism survey to provide a complete view of the properties, growth, and evolution of bright broad-line AGNs at $z>5$.

\end{abstract}

\keywords{high-redshift --- AGNs --- galaxies --- supermassive black holes}

\section{Introduction}
Over the past decades, studies of high-redshift supermassive black holes (SMBHs) have advanced significantly \citep{Fan2023}.  JWST has undoubtedly expanded the discovery space toward higher redshifts and lower luminosities, owing to its unprecedented infrared capabilities. During its few years of operation, JWST has discovered and confirmed hundreds of broad-line emitters at $z>4$, revealing the possible prevalence of AGNs up to $z\sim8$ \citep[e.g.,][]{ Ubler2023, Goulding2023, Harikane2023, Larson2023,Matthee2024, Kokorev2024, Greene2024, Maiolino2024, Taylor2024}. The broad Balmer emission from these objects, if originating from the broad-line region (BLRs) around SMBHs, suggests BH masses $<10^9M_\odot$, on average 1-2 dex lower than those of UV-luminous quasars at similar redshifts. In contrast, their number densities are 10-100$\times$ higher than the extrapolation from the quasar luminosity function \citep[e.g.,][]{Matthee2024, Greene2024, Lin2024, Maiolino2024}.  The growing discoveries of high-redshift AGNs have motivated new theoretical frameworks and simulations for BH seeding and growth in the early Universe \citep{Volonteri2023, Trinca2024, Jeon2025, SZhang2025, Volonteri2025, HZhang2025}.

Among the type-1 AGNs discovered by JWST, an enigmatic subsample, nicknamed ``little red dots'' (LRDs), has quickly become the focus of intense debate \citep[e.g.][]{Greene2024, Labbe2025}. The term LRD was originally used for broad-line selected AGNs in \citet{Matthee2024}, but soon came to specifically refer to objects with distinctive ``v-shaped'' SEDs that set them apart from typical AGNs \citep[e.g.,][]{Kocevski2024, Akins2024}.  These v-shaped objects are characterized by red optical continua and relatively blue UV continua, with the spectral turnover consistently located near the Balmer break \citep[e.g.,][]{Wang2024a, Setton2024, Labbe2025}. They also exhibit intriguing properties that were rarely seen prior to JWST, including prevalent Balmer absorption \citep{Lin2024, Juodzbalis2024}, {weak infrared and sub-millimeter emission \citep{Williams2024, Barro2024, Akins2024, Casey2024, Labbe2025, Setton2025, Li2025, Chen2025, Casey2025}, weak X-ray \citep{Yue2024, Ananna2024, Sacchi2025}, and weak radio emission \citep{Mazzolari2024, Perger2025}}. The current NIRCam grism surveys indicate that 10\%-20\% of high-$z$ broad-line AGNs are associated with \ha\ absorption, which is rarely observed among low-redshift type-1 AGNs \citep{Lin2024}.  Most of them do not exhibit significant photometric variability \citep{Kokubo2024, Zhang2024}. For example, \citet{Tee2024} report $\Delta m_{\rm UV} \approx 0.1 \pm 0.2$ mag for LRDs at $z=3-8$. New observations have triggered widespread discussion and inspired various hypotheses,  from super-Eddington accreting BHs \citep{Lambrides2024}, BHs embedded in dense gas \citep{Inayoshi2025, Ji2025, Naidu2025}, and even non-BH origins \citep{Baggen2024}. 

To understand BH populations in the early Universe and the enigmatic nature of LRDs within the broader AGN context, demographic studies based on large samples are crucial.   It is essential to conduct wide-area, multi-wavelength, and multi-epoch surveys to assemble a large sample of spectroscopically confirmed broad-line AGNs at $z>5$ and statistically study their properties. COSMOS-3D  (GO-5893; PIs Kakiichi, Egami, Fan, Lyu, Wang, and Yang) is a 268 hour JWST Cycle 3 Treasury program that mainly utilizes NIRCam \citep{rieke23a} Wide Field Slitless Spectroscopy (WFSS) over 0.33 deg$^2$ in the COSMOS field, with additional broad-band NIRCam and MIRI \citep{Wright23} imaging observations.  With its extensive coverage, COSMOS-3D is expected to yield a large sample of spectroscopically confirmed AGNs through a blind broad-line search. The COSMOS-3D AGN sample is expected to include a substantial number of $z=5-7$ broad \ha\ emitters and $z=7-9$ emitters with bright broad \hb\ emission. It will provide a large dataset to constrain the bright-end of the luminosity function of broad-line AGNs. The spectral resolution of WFSS ($R\sim1600$) is sufficient to resolve Balmer absorption, allowing COSMOS-3D to determine its prevalence at $z>5$ with a statistically robust sample. Additionally, COSMOS-3D includes NIRCam F115W observations overlapping with the COSMOS-Web program (GO-1727, PIs Kartaltepe and Casey), enabling a wide-area study of rest-frame UV variability at $z>5$. COSMOS-3D includes F1000W and F2100W MIRI parallel observations, making it the largest (482 arcmin$^2$) deep mid-infrared imaging survey at 10-20\,\micron\ currently. Combined with existing MIRI imaging in the COSMOS-Web field \citep{Harish2025}, COSMOS-3D will place strong constraints on the near-infrared SEDs of  AGNs up to rest-frame $\gtrsim3$\,\micron, and enable the search of high-$z$ obscured AGNs. The existing ALMA data  \citep{Casey2021, Long2024} and ongoing survey (e.g., CHAMPS, Faisst et al., in prep) will bring further insights into the sub-millimeter emission. COSMOS-3D will also enhance our understanding of the large-scale environmental impact on AGNs by mapping emitters at similar redshifts surrounding them. 

In this paper, we present the first results on broad-line AGNs at $z>5$ from COSMOS-3D. This first dataset, covering 127 arcmin$^2$ and representing just 10\% of the full survey, already demonstrates the capabilities of COSMOS-3D in studying the nature of high-redshift AGNs.  This work serves as a preview of the full potential of the survey. Detailed analyses of the complete COSMOS-3D sample will follow in future papers once the full dataset is available.

The paper is organized as follows. In \S\ref{sec:data_sample}, we introduce the datasets and methods to select broad-line AGNs. We present the analysis and results in \S\ref{sec:result}, including the photometric and spectroscopic properties, variability, and luminosity function. We discuss the abundance of $z>7$ AGNs in \S\ref{sec:discussion}. We adopt the AB magnitude system for all photometric measurements in this paper. Throughout this work, a flat $\Lambda$CDM cosmology is assumed, with $\rm H_0 = 70~km~s^{-1}~Mpc^{-1} $, $\Omega_{\Lambda,0} = 0.7$, and  $\Omega_{m,0}=0.3$.

\section{Data and sample}\label{sec:data_sample}

\subsection{JWST/NIRCam  Grism spectroscopy}
The COSMOS-3D program aims to conduct JWST/NIRCam WFSS observation in the F444W filter with Grism R over a 0.33 deg$^2$ area in the COSMOS-Web footprint \citep{Casey2023}. As of 2024 December 21, 10\% of the program has been completed. This first dataset includes two observations covering a total area of approximately 127 arcmin$^2$. Each observation contains mosaics arranged in four rows and two columns, and each pointing has an exposure time of 1868\,s. The data reduction and spectral extraction are detailed by Wang et al.\ (in prep.)  Here, we briefly
describe the procedure. For individual NIRCam WFSS exposures, we correct for the $1/f$ noise, apply flat-fielding, and subtract from all current COSMOS-3D WFSS observations the sky background using median background models. We align the WCS frames and measure the astrometric offsets between the short-wavelength images (taken with the grism observations) and the COSMOS-Web catalogs \citep{Shuntov2025}. These offsets are then applied to the spectral tracing model. Spectral tracing follows the procedure of \cite{Sun2023}. The spectral tracing and wavelength calibration are based on the Commissioning, Cycle-1, and Cycle-2 calibration data from the SMP-LMC-58 field, collected by 2024 June (PID 1076, 1479, 1480, and 4449; see the most recent update by \citealt{Sun2025})\footnote{\url{https://github.com/fengwusun/nircam_grism}}. The flux calibration is based on the Cycle-1 calibration data (PID: 1076, 1536, 1537, 1538).  We stack the 2D spectra of each object and optimally extract their 1D spectra following \cite{Horne1986}. In addition, we construct the line spectra by subtracting the median-filtered continuum models from the 2D spectra and then optimally extracting them. To identify emission lines, we apply a peak-finding algorithm. We refer to \cite{Wang2023} for more details on the emission line searching algorithm. The 5$\sigma$ line flux limits, based on a 5 pixel high boxcar extraction and integrated over a 300 \si{km\,s^{-1}} velocity width, are $4\times10^{-18}$\,\si{erg\,s^{-1}\,cm^{-2}} at 4\,\micron\ and $6\times10^{-18}$\,\si{erg\,s^{-1}\,cm^{-2}} at 4.9\,\micron, respectively.

\subsection{NIRCam imaging and photometric catalog}\label{sec:image_reduction}
The COSMOS-3D program acquires deep F200W images simultaneously with the WFSS observations. It also obtains F115W and F356W images, each with a 933s exposure time per pointing. Together with the COSMOS-Web observations, the current COSMOS-3D field now includes F115W, F150W, F200W, F277W, F356W, and F444W imaging. The images are processed following the procedures outlined by \cite{Yang2023}, and the photometric catalog is constructed as described by Champagne et al. (2025, in prep). We use the JWST pipeline 1.17.0 and calibration reference files \texttt{jwst\_1321.pmap}. In addition to the default steps in the pipeline for processing the raw data, we apply custom steps to remove the $1/f$ noise. The sky background is subtracted using master background models constructed from all the current COSMOS-3D images in each band. The WCS frame is aligned to match that of the COSMOS-Web catalog \citep{Shuntov2025}.

To construct the photometric catalog, we start with a $\chi^2$ stacked detection image that combines all available JWST data in the field (F115W, F150W, F277W, and F444W from COSMOS-Web, along with F115W, F200W, and F356W from COSMOS-3D).
Source detection is performed using \texttt{SourcExtractor++} \citep{Bertin2020} with a hot+cold detection scheme optimized to capture both large, bright sources and small, faint ones (following a similar procedure to \citealt{Shuntov2025} for COSMOS-Web).  
Photometry is measured in 0.32 arcsec circular apertures. We apply aperture corrections to all filters, defined as the ratio of the flux in a default Kron aperture ($k$=2.5, $R_{\min}$=3.5) to that in a smaller Kron aperture ($k$=1.2, $R_{\min}$=1.7).
We measure the noise by placing 1000 random apertures across each image with detected sources masked. We then calculate the standard deviation as a function of aperture size, which is added in quadrature with the Poisson noise from the ERR map. All photometry is corrected for Galactic extinction.

\subsection{Ancillary data}
For the selected sample (see \S\ref{sec:sample_selection}), we retrieve 30\arcsec$\times$30\arcsec\ Hubble Space Telescope (HST) Advanced Camera for Surveys (ACS) F814W cutouts from the COSMOS public data release\footnote{\url{https://cosmos.astro.caltech.edu/page/hst}} \citep{Scoville2007, Koekemoer2007}, align the WCS to our reference frame, and measure the photometry using the same \texttt{KRON} parameters as for the NIRCam catalog. 

The current COSMOS-3D grism footprint partially overlaps with the MIRI F770W and F1800W imaging from COSMOS-Web and the PRIMER program (GO 1837, PI Dunlop). We retrieve MIRI data from the DAWN JWST Archive\footnote{\url{https://dawn-cph.github.io/dja/index.html}} \citep{grizli, Valentino2023}. For the subset of our sample with MIRI coverage, we measure photometry using $0\farcs5$
 circular apertures. Aperture corrections are applied based on empirical point-spread functions constructed by \cite{Alberts2024}, and photometric uncertainties are estimated from the scatter of 500 randomly placed apertures near each source.

We further match the sources to the COSMOS2020 catalog \citep{Weaver2022} using a $0\farcs5$ matching radius. The COSMOS2020 catalog provides multiwavelength photometry, including HSC broad bands ($g, r, i, z, y$) and Subaru intermediate and narrow bands (ranging from IB427 to NB816).

\subsection{Sample selection}\label{sec:sample_selection}

To select $z>5$ broad-line emitters, we first systematically search for compact emitters based on the following criteria:
\begin{itemize}
    \item[(1)] Half-light radius $<$ $0\farcs25$;
    \item[(2)]  F444W flux ($<0\farcs3$) / flux ($<0\farcs15$) $<$ 1.5;
    \item[(3)] F150W -- F444W $>$ 0.1 mag  or F200W -- F444W $>0.1$ mag;
    \item[(4)] Emission lines with integrated line flux signal-to-noise ratio (S/N) $>$ 2 when detected by the line searching algorithm.
\end{itemize}
The first and second criteria require the object to be compact. The third criterion can effectively exclude most low-redshift Paschen line emitters, whose \ha\ or [\ion{O}{3}] emission lines fall within the F150W or F200W filters.  In the fourth criterion, we apply a low S/N cut of 2 to ensure an inclusive selection. This is because the broad wings of the lines can be oversubtracted during median filtering, potentially leading to an underestimated S/N by the automatic line searching algorithm.

We perform the following steps to clean the parent sample selected from the criteria mentioned above. First, we visually inspect the spectra to identify fake sources, false line detections, and spectra contaminated by emission from other sources. Then, we exclude low-redshift emitters with photo-$z < 3.5$ that show Paschen $\alpha$, Paschen $\beta$, [\ion{S}{3}]\,$\lambda\lambda$9071,9533, or \ion{He}{1}\,$\lambda$10833. Given the depth of COSMOS-3D, we find that robust broad-line detection is only possible with bright sources. As a result, we primarily focus on emitters with F444W $<$ 26 mag. We have confirmed that no reliable broad lines can be detected in fainter sources based on the line-fitting criteria outlined below. We end up with 28 compact \ha\ and \hb+[\ion{O}{3}] emitters with F444W $<$ 26 mag.

We identify broad Balmer components by fitting the \ha\ and \hb\ emission line profiles. We adopt two models: first with a single Gaussian model, and then with a two-Gaussian model. The criteria to select broad-line emitters are:
\begin{itemize}
    \item[(5)] At least one of the Gaussian profiles exhibits a full width at half maximum (FWHM) greater than 1000 km s$^{-1}$;
    \item[(6)] The broad component  has an integrated line flux S/N $>$ 5.
\end{itemize}

This process identifies 13 broad-line emitters, including 12 broad \ha\ emitters at $z=5-6$ and 1 broad \hb\ emitter at $z>7$.  The wavelength range of the F444W grism can ideally capture emitters at $z = 4.9 - 9.0$. However, in practice, the redshift range of the broad-line emitters is influenced by the wavelength-dependent sensitivity of the filters (\citealt{rieke23a}). The detected \ha\ and \hb\ broad lines are observed in the wavelength range 3.95--4.60\,\micron, where the F444W grism is the most sensitive.

\begin{figure}
    \centering
    \includegraphics[width=\linewidth]{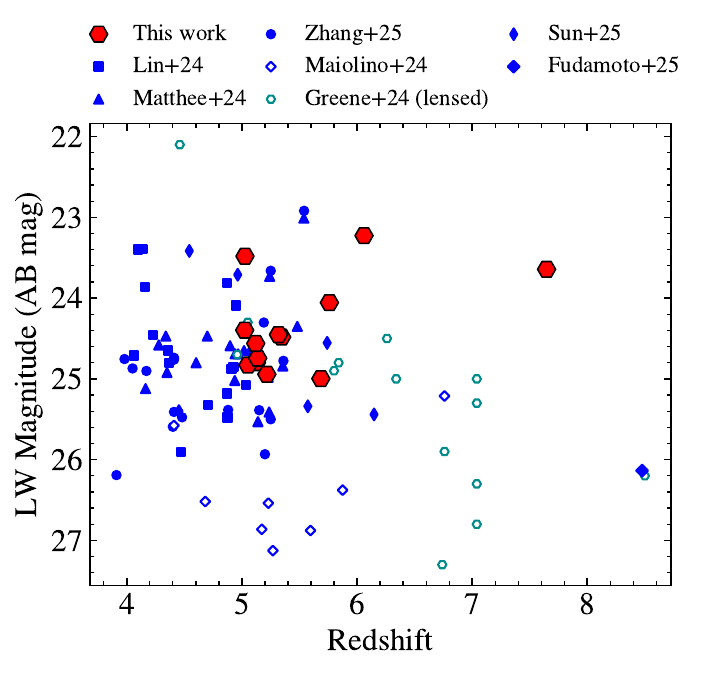}
    \caption{Long-wavelength magnitude (LW Magnitude) as a function of redshift for the broad-line AGN sample in this work, compared to literature samples \citep{Matthee2024,Maiolino2024, Greene2024, Lin2024, Zhang2025, Sun2025, Fudamoto2025}. Filled markers represent broad-line AGNs selected from NIRCam/WFSS, and open markers denote those identified with NIRSpec. Note that the sample from \cite{Greene2024} has not been corrected for lensing magnification. LW magnitude refers to either F356W for $z<5$ samples or F444W for $z>5$ samples. }
    \label{fig:z_magnitude}
\end{figure}

We present the redshift versus F444W magnitude of our target sample in Figure \ref{fig:z_magnitude}. These broad-line emitters are compared to existing samples from the literature. The broad-line emitters in this work are at the bright end compared to previous samples, with a median LW (F356W or F444W) magnitude of 24.5 mag. We have four objects with F444W magnitudes brighter than 24 mag. Among them, we have an object with an F444W magnitude of 23.64 at $z = 7.646$. This is the brightest JWST-discovered broad-line AGN at $z > 7$ to date. We name it as \hbagn. In comparison, before JWST, the highest redshift UV-luminous quasar known to date \citep{Wang2021} is at $z=7.642$ with a Wide-field Infrared Survey Explorer W2 magnitude of 20.2 $\pm$ 0.3, and GNz7q at $z=7.19$ \citep{Fujimoto2022} has an IRAC 4.5-\micron\ magnitude of 22.4 $\pm$ 0.1.

\subsection{Comparison with photometric selected LRDs}
\begin{figure}[t!]
    \centering
    \includegraphics[width=\linewidth]{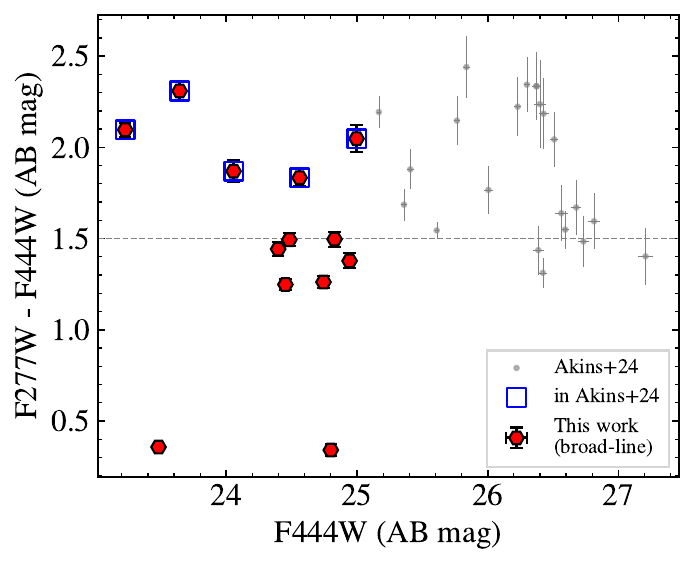}
    \caption{The broad-line AGNs in this work are plotted in the F277W - F444W vs. F444W color space. The gray dots represent the photometrically selected LRDs from \cite{Akins2024} within the current COSMOS-3D footprint, but with photometry measurements taken from our catalog (\S\ref{sec:image_reduction}). The dashed lines indicate the selection criteria for these LRDs. The broad-line AGNs in this work are selected based on the presence of broad emission lines and, therefore, do not necessarily satisfy the LRD color selection criteria. 
    }
    \label{fig:F277W_F444W}
\end{figure}

We compare our sample to the photometrically selected LRDs from \cite{Akins2024} in Figure \ref{fig:F277W_F444W}. There are 27 sources in \cite{Akins2024} that fall within the current COSMOS-3D grism footprint. Based on the photo-$z$ reported by \cite{Akins2024}, 8 sources should have [\ion{O}{3}] $\lambda$5008 detected in the F444W grism, and for 18 sources the \ha\ emission line should have been detected. Among the 26 objects, emission lines with integrated flux S/N$>$5 are detected in 10 sources.  Two of these display [\ion{O}{3}] $\lambda\lambda$4960,5008 doublets, and eight  show single emission lines which could correspond to either \ha\ or [\ion{O}{3}] $\lambda$5008. Among them, five objects are in our sample: four broad-line \ha\ emitters and one broad-line \hb\ emitter. We note that the detectability of broad lines may be limited by the shallow grism exposures, where the faint broad components could be overwhelmed by the high zodiacal background. Also, for the sources without detected lines, the expected \ha\ or [\ion{O}{3}] emission may fall outside the F444W grism coverage due to the uncertainties of photo-$z$. Overall, the LRD sample of \cite{Akins2024} shows high purity for sources with F444W$<$25\,mag. 

These five overlapping sources represent only 38\% of our sample, while the remaining sources do not satisfy the LRD selection in \cite{Akins2024}. This is because \cite{Akins2024} prioritize sources with extremely red optical colors to achieve high purity in their sample and to minimize contamination from extreme emission lines. In contrast, our selection is based solely on a broad-line search without constraints on the SED shape, allowing us to include sources with bluer optical continua. LRDs represent only a small subset of the high-redshift AGN population \citep{Hainline2025}.

\section{Results}\label{sec:result}

\subsection{Broad-line profile}\label{sec:line_profile}
We perform a more precise line profile fitting for the 13 selected broad-line emitters, taking the line spread functions (LSFs) into account. The construction of NIRCam WFSS LSFs will be described by Sun et al. (in prep.; see also \citealt{Danhaive25}). Based on the observed line profiles of these broad-line emitters, we adopt different assumptions for their intrinsic line profiles and, thus, different fitting procedures. All line profile fitting is performed on the original optimally extracted spectra without continuum removal. 

\begin{figure*}
    \centering
    \includegraphics[width=0.49\linewidth]{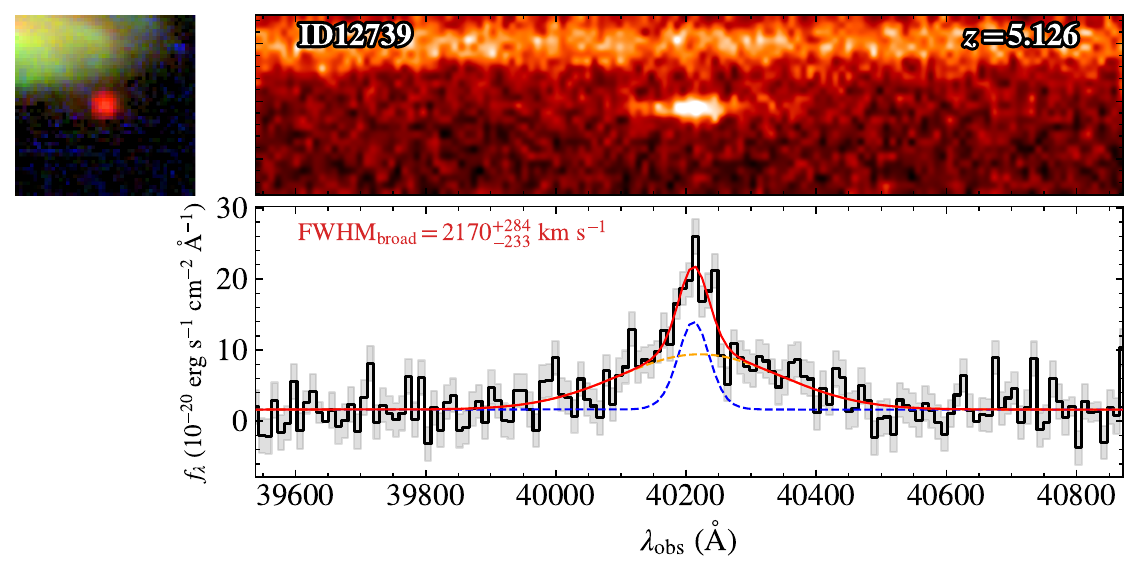}
    \includegraphics[width=0.49\linewidth]{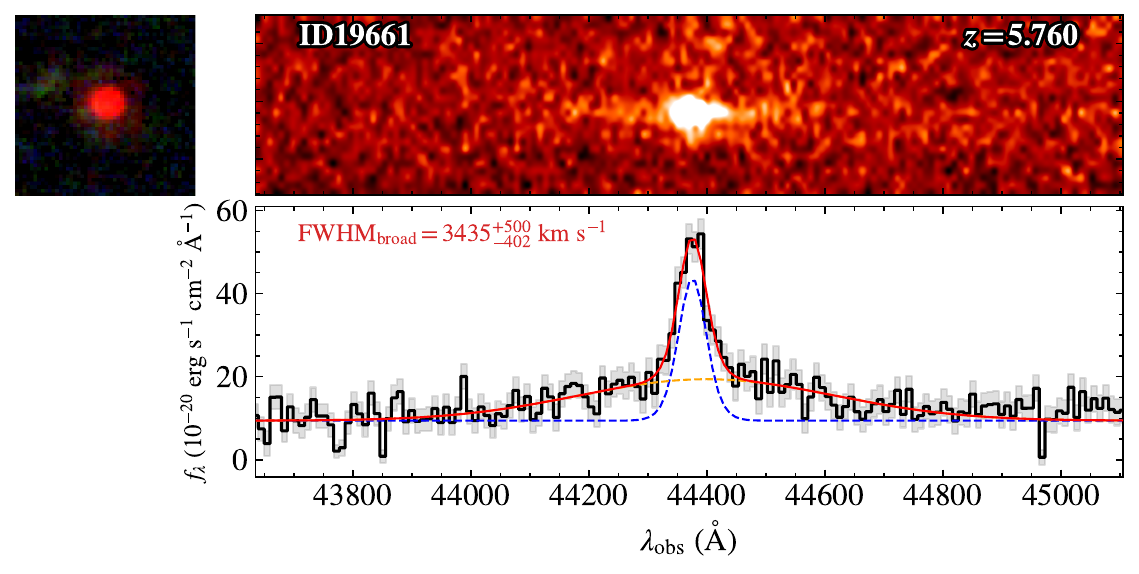}
    \includegraphics[width=0.49\linewidth]{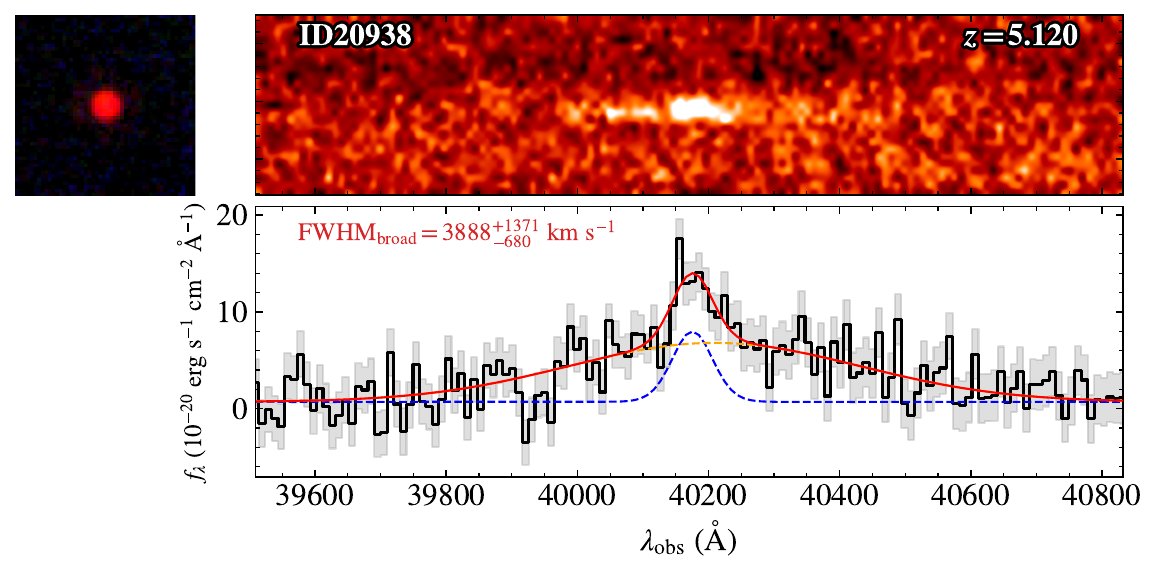}
    \includegraphics[width=0.49\linewidth]{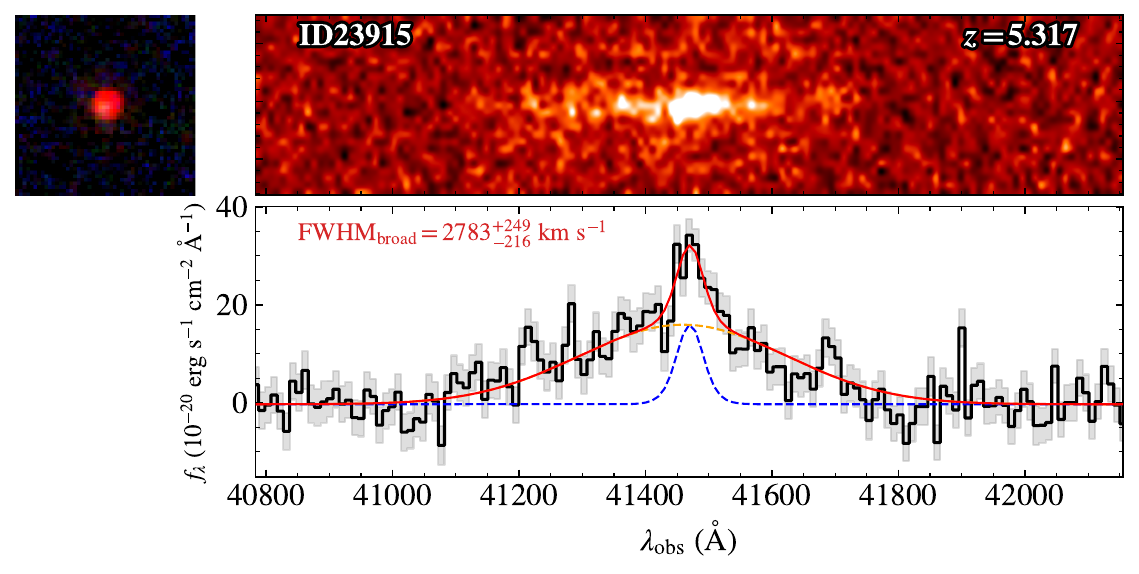}
    \includegraphics[width=0.49\linewidth]{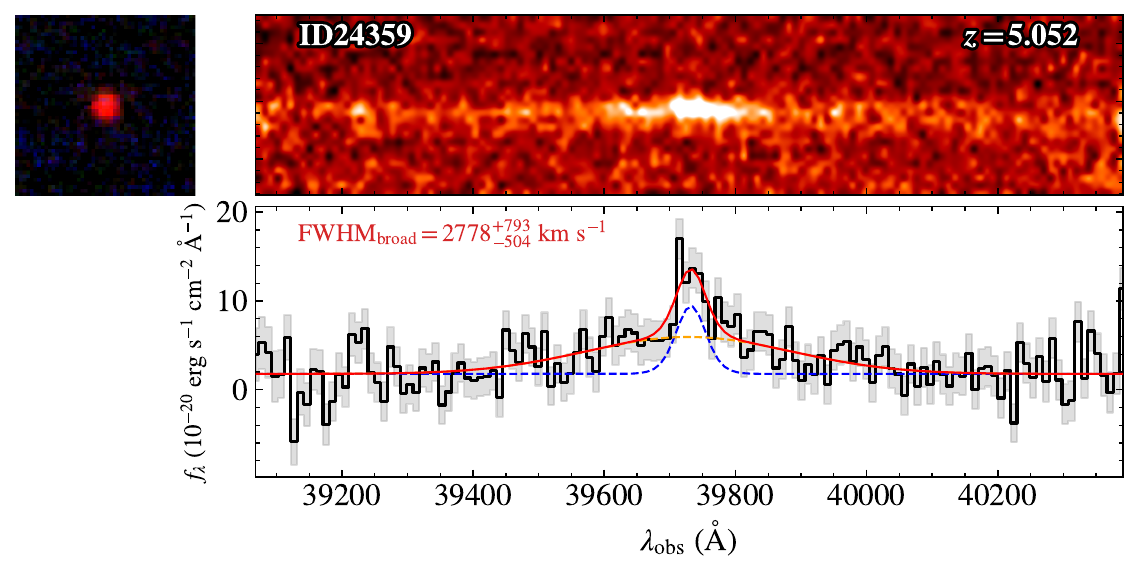}
    \includegraphics[width=0.49\linewidth]{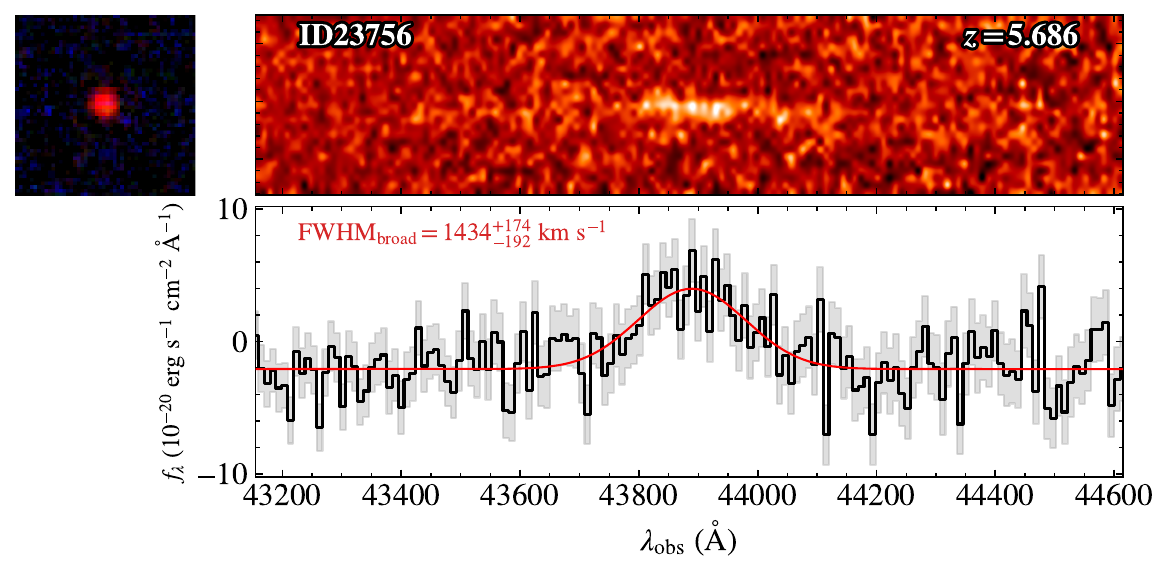}
    \includegraphics[width=0.49\linewidth]{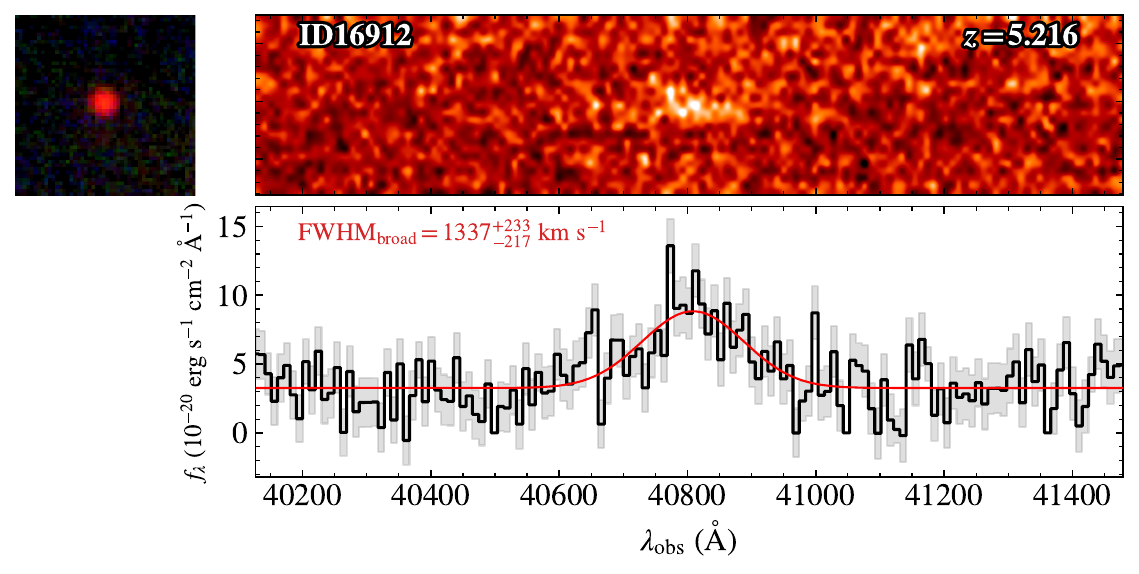}
    \includegraphics[width=0.49\linewidth]{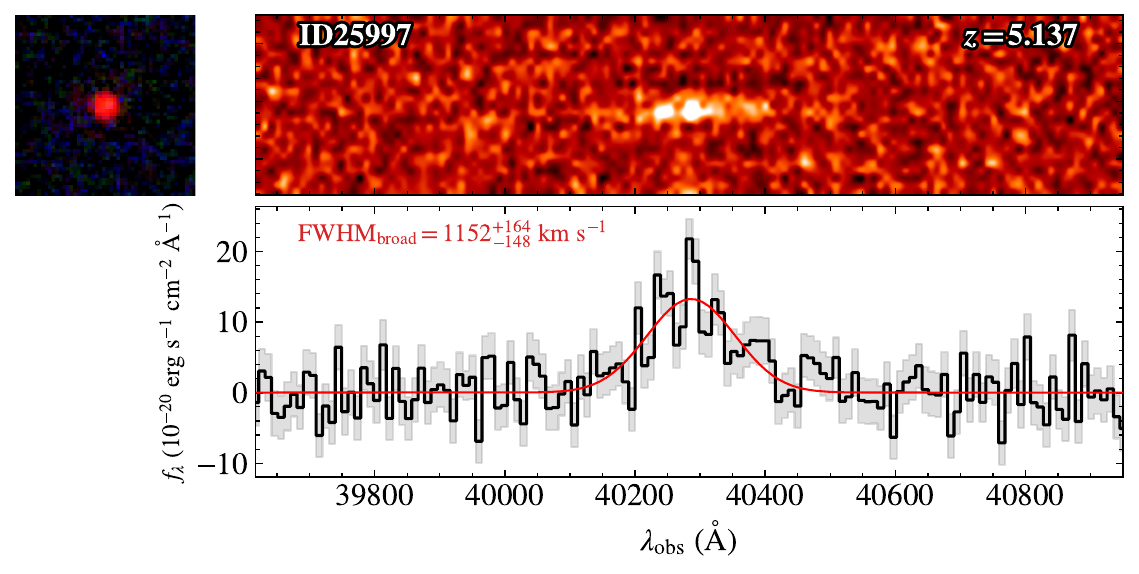}
    \caption{Eight broad-\ha\ emitters in the first 10\% COSMOS-3D footprint. For each source, the upper-left panel displays a 2\arcsec$\times$2\arcsec\ RGB thumbnail composed of \textit{JWST}/NIRCam F444W, F277W, and F115W images. The top panel shows the 2D grism spectrum and the bottom panel presents the optimally extracted 1D spectrum (black) along with the corresponding error spectrum (gray-shaded region). The best-fit line profiles are shown as a solid red line.  If the profiles include two Gaussian components, the narrow component is depicted as a blue dashed line and the broad component as an orange dashed line. All models have been convolved with the LSFs. The FWHMs of the broad emission lines are labeled for each object. }
    \label{fig:GrismHa}
\end{figure*}

\begin{figure*}
    \centering
    \includegraphics[width=0.48\linewidth]{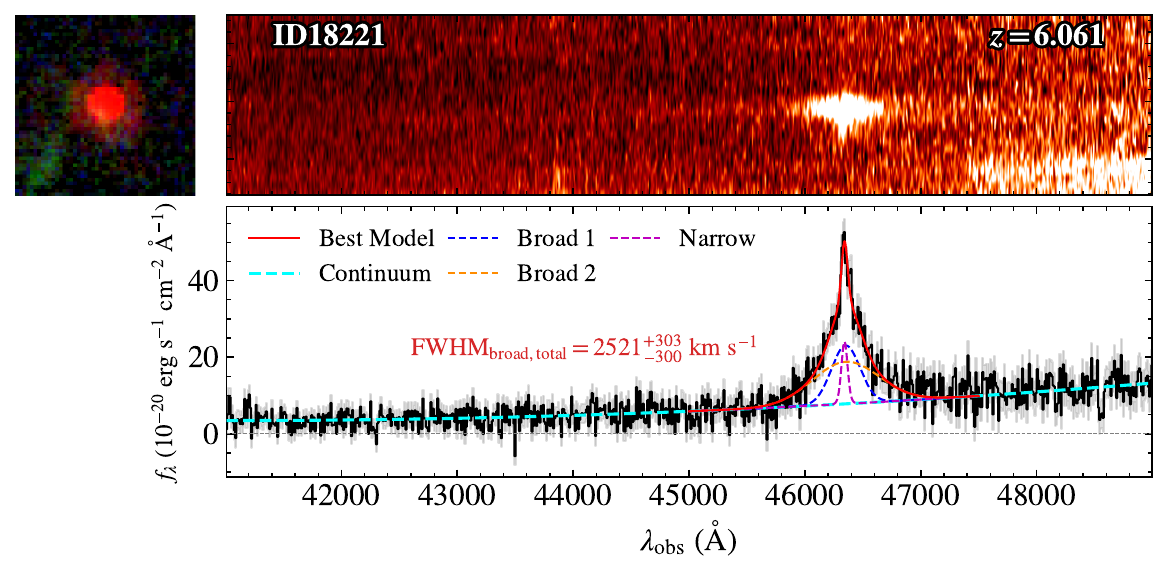}
\includegraphics[width=0.5\linewidth]{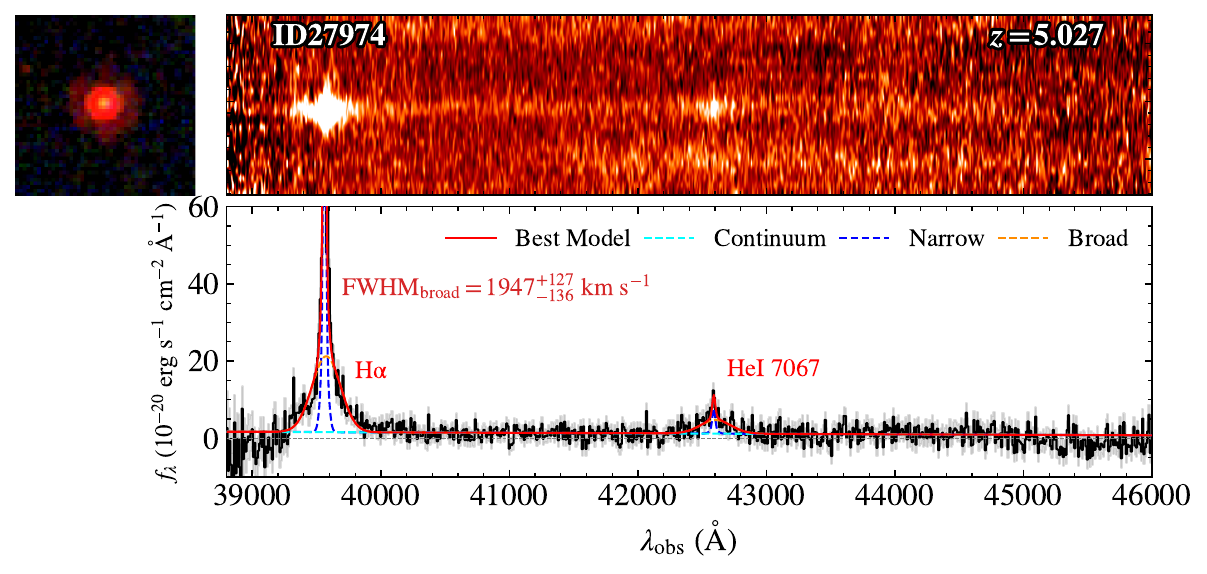}
    \caption{Similar to Figure \ref{fig:GrismHa}, but showing the spectra of ID18221 and ID27974, which display continuum emission. } 
    \label{fig:GrismHa_withContinuum}
\end{figure*}

\begin{figure*}
    \centering
    \includegraphics[width=0.49\linewidth]{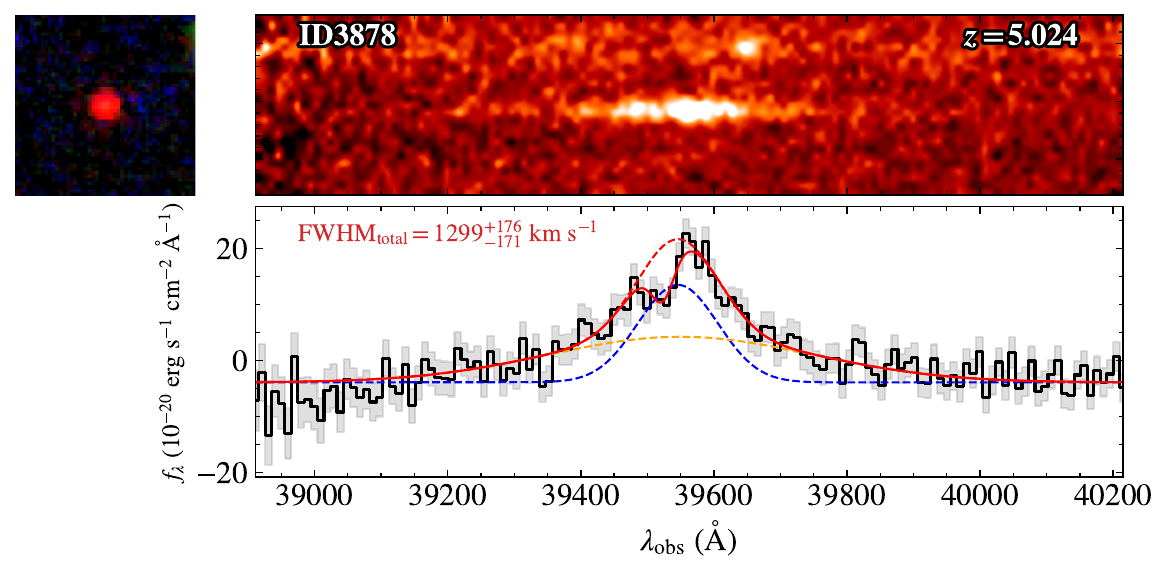}
    \includegraphics[width=0.49\linewidth]{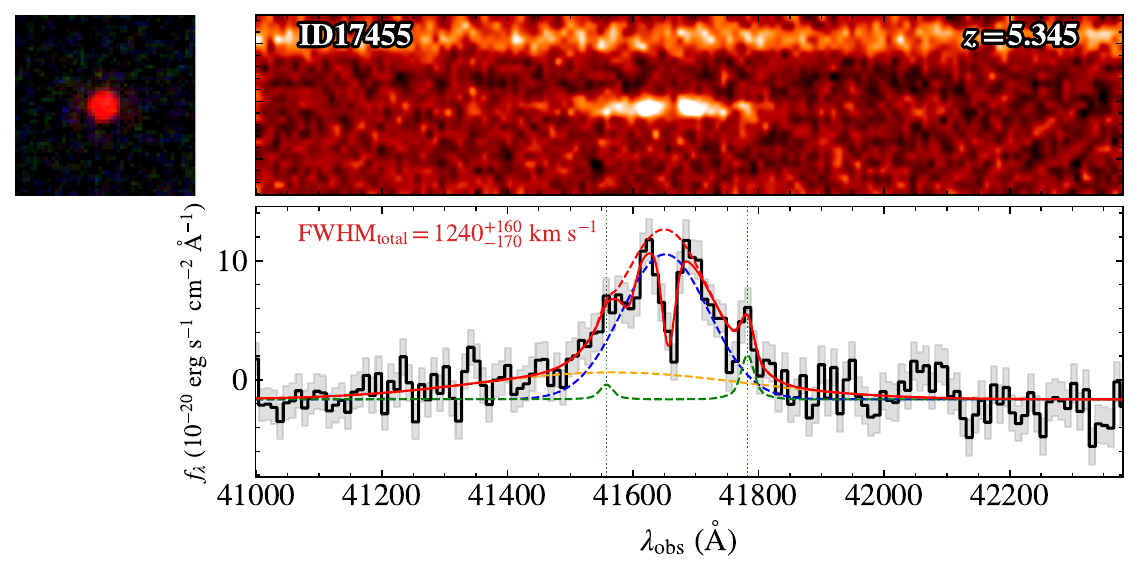}

    \caption{Similar to Figure \ref{fig:GrismHa}, but showing the spectra of ID3878 and ID17455 with \ha\ absorption. The intrinsic profiles without absorption are shown as red dashed lines.  For ID17455, its [\ion{N}{2}] emission is shown as a green dashed line.   }
    \label{fig:GrismHa_abs}
\end{figure*}

\begin{figure*}
    \centering
    \includegraphics[width=\linewidth]{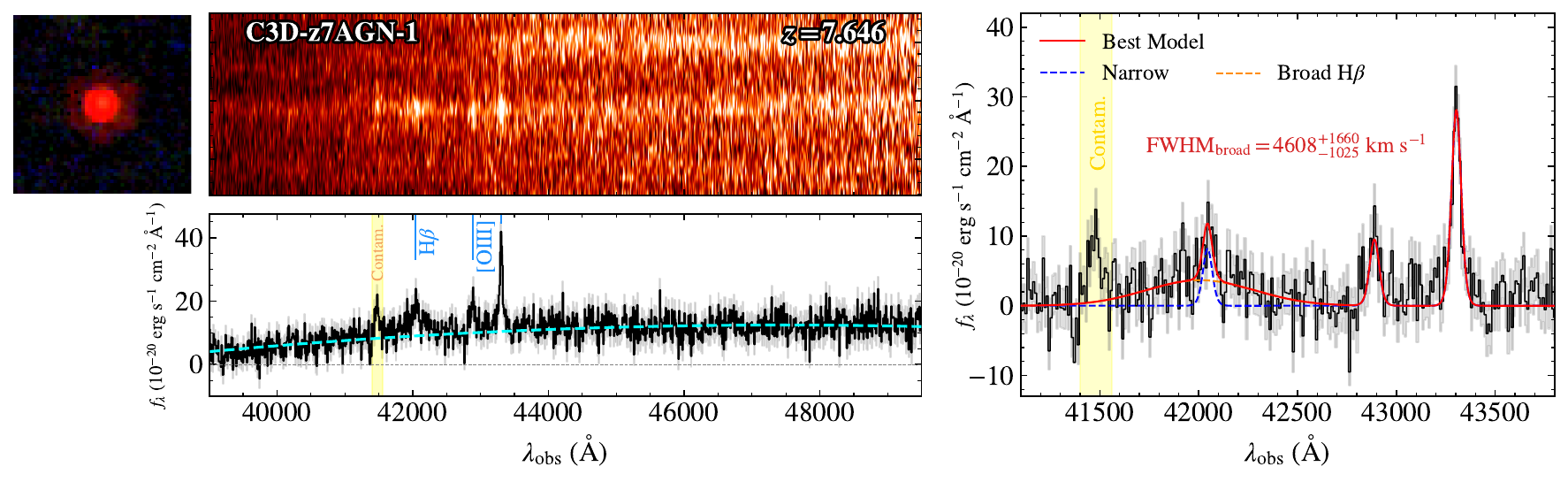}
    \caption{ \textit{Left}: Similar to Figure \ref{fig:GrismHa}, but showing the spectra of C3D-z7AGN-1 (ID13852), which exhibit broad \hb\ emission lines. The cyan dashed line represents the continuum. The contaminated emission from other sources is masked, indicated by the yellow-shaded region. \textit{Right}: continuum-subtracted spectra around \hb\ and [\ion{O}{3}] lines.
    \label{fig:GrismHbO3}}
\end{figure*}

\begin{itemize}
    \item  \textbf{Narrow+broad \ha}. For five broad-line emitters (ID12739, ID19661, ID20938, ID23915, and ID24359), their line profiles can be well described by a combination of narrow  and broad Gaussian profiles, with FWHMs of the latter exceeding 1000 km s$^{-1}$.  We also add a constant flux density offset to the model to represent the local continuum level.  We present their spectra in Figure \ref{fig:GrismHa}.
    
    ID27974 features a strong \ion{He}{1} $\lambda$7067 line and a detected continuum, as shown in Figure \ref{fig:GrismHa_withContinuum}. We assume Gaussian profiles for \ion{He}{1} with the same FWHMs as those of \ha\ but with different amplitudes. We use a linear model to represent the continuum between \ha\ and \ion{He}{1}, since this tentative continuum cannot be well modeled with a power law.
 
    ID18221 exhibits a significant continuum and a narrow component more extended than the point source, attributed to the host galaxy, as shown in Figure \ref{fig:GrismHa_withContinuum}. We fit and remove the continuum using a degree-two spline interpolation. The continuum-subtracted line profile fits with three Gaussian components: two broad Gaussians for the emission from the central point source and one narrow Gaussian for the emission from the host galaxy or AGN narrow-line region (NLR).
    
    \item \textbf{Broad \ha\ only}. For three emitters (ID23756, ID16912, and ID25997), limited by the S/N, we assume their profiles are dominated by broad Gaussian components with FWHM $>$ 1000 km s$^{-1}$.  We present their spectra in Figure \ref{fig:GrismHa}.

    \item  \textbf{\ha\ with absorption}. The line profiles of ID3878 and ID17455 show significant \ha\ absorption (Figure \ref{fig:GrismHa_abs}). For 3878, we model the intrinsic profile as a combination of narrow and broad Gaussian components and the absorber with a Gaussian profile applied over the intrinsic profile\footnote{The absorbed profiles = the intrinsic emission profiles * (1 - absorber profiles)}. 
    
    For ID17455, the intrinsic line profiles include two Gaussian models for the emission, two absorbers, and [\ion{N}{2}] emission lines. We assume the two absorbers and the [\ion{N}{2}] emission lines to be unresolved.    
    The absorption is superimposed on the Gaussian components of the \ha\ emission.

    For both emitters, a constant is added to account for the local continuum level, which could arise from either the residual from sky background subtraction or continuum contaminants from other sources. 
 
    \item \textbf{\hb+[\ion{O}{3}] with continuum}. \hbagn\ (ID13852) presents \hb+[\ion{O}{3}] emission with a continuum rising toward  longer wavelengths (Figure \ref{fig:GrismHbO3}). We use degree-two spline interpolation to model and subtract the continuum. We model the [\ion{O}{3}] doublets with a single Gaussian component and an amplitude ratio of 1:2.98. For \hb, we assume a narrow Gaussian component with the same FWHM as the [\ion{O}{3}] doublets and a separate broad Gaussian component. During the fitting, we mask the contamination from another source blueward of \hb.
\end{itemize}

We convolve the assumed line profile models with the LSFs and fit the observed lines using the Markov Chain Monte Carlo (MCMC) method\footnote{\url{https://github.com/dfm/emcee}} \citep{emcee}.  The FWHMs and luminosities of the broad emission lines are listed in Table \ref{tab:property}.  The properties of the \ha\ absorbers are summarized in Table \ref{tab:absorption}. For three objects (ID18221, 3878, and 17455) with emission lines composed of two FWHM$>1000$\,\si{km\,s^{-1}} components, we calculate the FWHM and luminosity from the composite profile of the two broad components. We estimate the BH mass (\MBH) using the measured broad profiles following \cite{Greene2005}: 

\begin{equation}
\begin{aligned}
& M_{\mathrm{BH}}=\left(2.0_{-0.3}^{+0.4}\right) \times 10^6 \\
& \left(\frac{L_{\rm H\alpha, broad}}{10^{42} \mathrm{erg} \mathrm{~s}^{-1}}\right)^{0.55 \pm 0.02}\left(\frac{\mathrm{FWHM}_{\rm H\alpha, broad}}{10^3 \mathrm{~km} \mathrm{~s}^{-1}}\right)^{2.06 \pm 0.06} M_{\odot};
\end{aligned}
\end{equation}

\begin{equation}
\begin{aligned}
& M_{\mathrm{BH}}=(3.6 \pm 0.2) \times 10^6 \\
& \quad\left(\frac{L_{\rm H\beta, broad}}{10^{42} \mathrm{erg} \mathrm{~s}^{-1}}\right)^{0.56 \pm 0.02}\left(\frac{\mathrm{FWHM}_{\rm H\beta, broad}}{10^3 \mathrm{~km} \mathrm{~s}^{-1}}\right)^2 M_{\odot}.
\end{aligned}
\end{equation}

We present our \MBH\ measurements in Table \ref{tab:property}. Although we derive these properties following common practices in high-redshift AGN studies, we caution the potential systemic errors associated with these estimates. First, it is unclear if the BH mass calibration in \cite{Greene2005}, based on local type-1 AGNs, can be applied to high-redshift AGNs with significantly different SED shapes. Recent models and observations suggest dense gas surrounding the BHs \citep{Inayoshi2025,Ji2025, Naidu2025, deGraaff2025, Rusakov2025}. In this case, electron scattering may broaden the lines, and the values derived from the observed FWHMs may overestimate the BH masses by 1-2 dex \citep{Rusakov2025}.  Second, the origin of the narrow components in most of these AGNs — whether from the NLRs or the host galaxy — remains uncertain. Furthermore, the dust attenuation levels and laws in these objects are unknown. Therefore, $L_{\rm H\alpha, broad}$ should be considered a lower limit for the line luminosity originating from AGNs.

\begin{table*}
	\begin{center}
 
		\begin{tabular}{cccccccccc}
            \hline
			ID & RA & Dec & $z$ & F444W mag & FWHM$_{\rm broad}$ & $L_{\rm broad}$ & $\log M_{\rm BH}$ & $M_{\rm UV}$ & $\beta_{\rm opt}$ \\
			 &  &  &  &  & (km s$^{-1}$) & (10$^{42}$ erg s$^{-1}$) & ($M_\odot$) &  &  \\
            \hline
			3878 & 150.1634 & 2.4392 & 5.024 & $24.40\pm0.01$ & $1299^{+176}_{-171}$ & $17.74^{+1.83}_{-2.63}$ & $7.21^{+0.26}_{-0.25}$ & $>-18.78$ & $0.50 \pm 0.16$ \\
			27974 & 150.0545 & 2.5261 & 5.027 & $23.48\pm0.01$ & $1947^{+127}_{-136}$ & $14.34^{+0.63}_{-0.73}$ & $7.53^{+0.19}_{-0.18}$ & $-20.77\pm0.08$ & $-1.69 \pm 0.05$ \\
			24359 & 149.9233 & 2.5396 & 5.052 & $24.83\pm0.01$ & $2778^{+795}_{-504}$ & $4.38^{+1.03}_{-0.89}$ & $7.58^{+0.39}_{-0.35}$ & $>-18.24$ & $1.32 \pm 0.19$ \\
			20938 & 150.0664 & 2.4532 & 5.120 & $24.56\pm0.01$ & $3888^{+1372}_{-680}$ & $9.34^{+3.62}_{-1.78}$ & $8.09^{+0.45}_{-0.38}$ & $>-18.55$ & $1.51 \pm 0.24$ \\
			12739 & 149.9793 & 2.4981 & 5.126 & $24.80\pm0.02$ & $2170^{+284}_{-233}$ & $6.71^{+0.74}_{-0.75}$ & $7.45^{+0.25}_{-0.24}$ & $-19.81\pm0.23$ & $-0.78 \pm 0.16$ \\
			25997 & 149.9107 & 2.5351 & 5.137 & $24.74\pm0.01$ & $1152^{+164}_{-148}$ & $6.26^{+0.70}_{-0.65}$ & $6.86^{+0.25}_{-0.23}$ & $-18.91\pm0.20$ & $0.93 \pm 0.15$ \\
			16912 & 150.1104 & 2.5043 & 5.216 & $24.94\pm0.01$ & $1337^{+233}_{-217}$ & $3.19^{+0.44}_{-0.42}$ & $6.83^{+0.28}_{-0.27}$ & $-18.59\pm0.28$ & $0.88 \pm 0.09$ \\
			23915 & 149.9034 & 2.4859 & 5.317 & $24.45\pm0.01$ & $2783^{+249}_{-216}$ & $20.13^{+1.74}_{-1.68}$ & $7.94^{+0.23}_{-0.21}$ & $-19.32\pm0.18$ & $-0.97 \pm 0.10$ \\
			17455 & 149.9729 & 2.5549 & 5.345 & $24.48\pm0.01$ & $1240^{+160}_{-170}$ & $9.63^{+1.54}_{-1.25}$ & $7.03^{+0.26}_{-0.25}$ & $>-18.67$ & $1.42 \pm 0.22$ \\
			23756 & 150.0667 & 2.4897 & 5.686 & $25.00\pm0.01$ & $1434^{+174}_{-192}$ & $4.86^{+0.64}_{-0.61}$ & $6.99^{+0.25}_{-0.24}$ & $-18.46\pm0.30$ & $1.50 \pm 0.13$ \\
			19661 & 150.0537 & 2.4058 & 5.760 & $24.06\pm0.01$ & $3435^{+500}_{-402}$ & $19.66^{+2.82}_{-2.27}$ & $8.12^{+0.29}_{-0.27}$ & $-19.62\pm0.30$ & $0.14 \pm 0.10$ \\
			18221 & 150.0586 & 2.3993 & 6.061 & $23.23\pm0.01$ & $2521^{+303}_{-300}$ & $51.41^{+2.19}_{-2.31}$ & $8.07^{+0.26}_{-0.25}$ & $-20.51\pm0.18$ & $0.96 \pm 0.08$ \\
			13852$^{\star}$ & 149.9514 & 2.4488 & 7.646 & $23.64\pm0.01$ & $4608^{+1660}_{-1025}$ & $17.69^{+3.53}_{-3.19}$ & $8.60^{+0.35}_{-0.33}$ & $-20.33\pm0.20$ & $1.12 \pm 0.05$ \\
        \hline
		\end{tabular}
        \caption{Broad \ha\ and \hb\ emitters in this work. For ID18221, ID3878, and ID17455, these values are derived from the composite profiles of the two broad Gaussian components.   For sources having rest-frame UV photometry with S/N $< 3$, we report the $3\sigma$ lower limits on $M_{\rm UV}$. $^{\star}$ID13852 is C3D-z7AGN-1. Its $\rm FWHM_{broad}$, $L_{\rm broad}$, and \MBH\ are measured from its broad \hb\ emission. 
         }
        \label{tab:property}

	\end{center}
\end{table*}

\begin{table}
	\begin{center}
		\begin{tabular}{cccc}
        \hline
			ID & $v_{\rm abs}$ & FWHM$_{\rm abs}$ & EW$_{\rm abs, 0}$ \\
			 & (\si{km\,s^{-1}}) & (\si{km\,s^{-1}}) & (\AA) \\
             \hline
			3878 & $-174^{+38}_{-48}$ & $214^{+57}_{-69}$ & $2.9^{+0.8}_{-0.8}$ \\
			17455 & $36^{+39}_{-35}$ & -- & $3.0^{+0.4}_{-0.5}$ \\
			17455 & $-417^{+80}_{-767}$ & -- & $1.2^{+0.7}_{-0.7}$ \\
            \hline
		\end{tabular}
        \caption{Properties of the Balmer absorption for the two sources. $v_{\rm abs}$ denotes the velocity of the absorption, FWHM$_{\rm abs}$ represents the full width at half maximum of the absorption, and EW$_{\rm abs, 0}$ is the rest-frame equivalent width of the absorption. The uncertainties are obtained from the MCMC fitting. The Balmer absorption in ID17455 is unresolved, so no FWHM value is provided.}
        \label{tab:absorption}
	\end{center}
\end{table}

\subsection{Multiwavelength SEDs}\label{sec:SED}

\begin{figure*}
    \centering
    \includegraphics[width=0.49\linewidth]{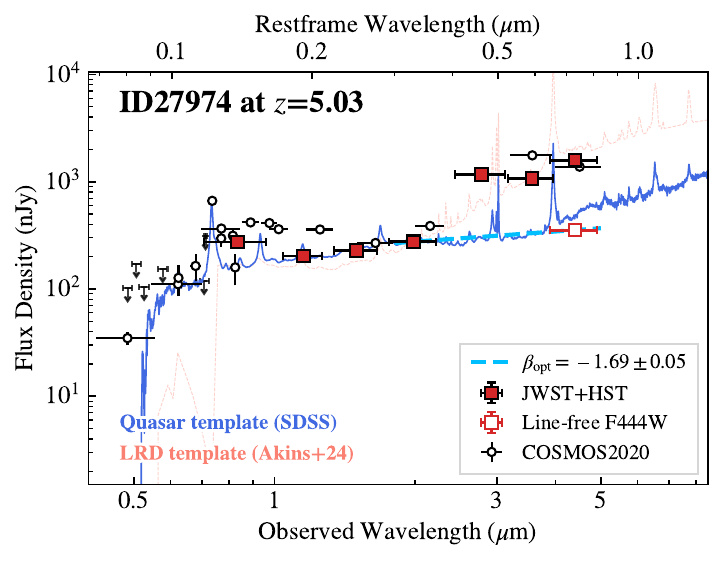}
    \includegraphics[width=0.49\linewidth]{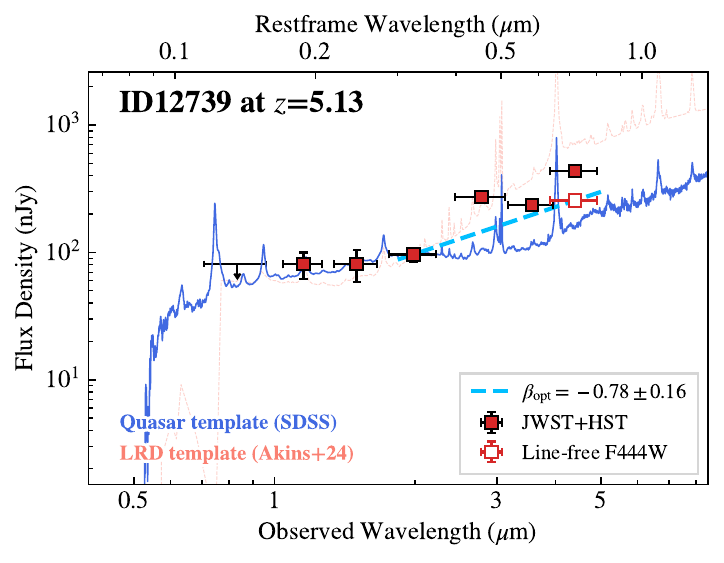}
    \caption{SEDs of ID27974 and ID12739 with blue optical continuum slopes ($\beta_{\rm opt} < 0$). The red squares represent photometry from HST/ACS F814W and JWST/NIRCam F115W, F150W, F200W, F277W, F356W, and F444W, while the red-edged white squares correspond to F444W photometry with the emission line subtracted. If the source is present in the COSMOS2020 catalog, the white circles represent its photometry from COSMOS2020. For photometry with S/N$<3$, we show the $3\sigma$ upper limits. The measured optical slopes are shown as blue dashed lines. We show the quasar template from \cite{VandenBerk2001} for reference (blue lines), and the LRD template from \cite{Akins2024} for comparison (red dashed lines). The templates are normalized to match the F200W magnitude.}
    \label{fig:quasar_like_SED}
\end{figure*}

\begin{figure*}
    \centering
    \includegraphics[width=0.49\linewidth]{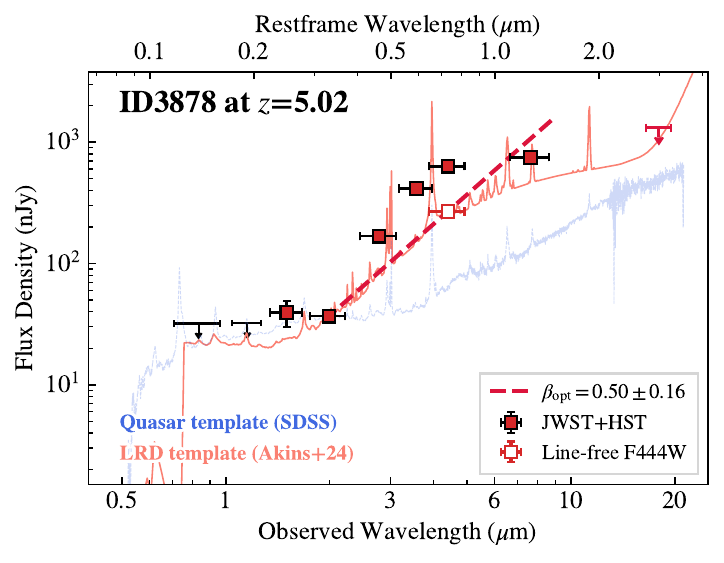}
    \includegraphics[width=0.49\linewidth]{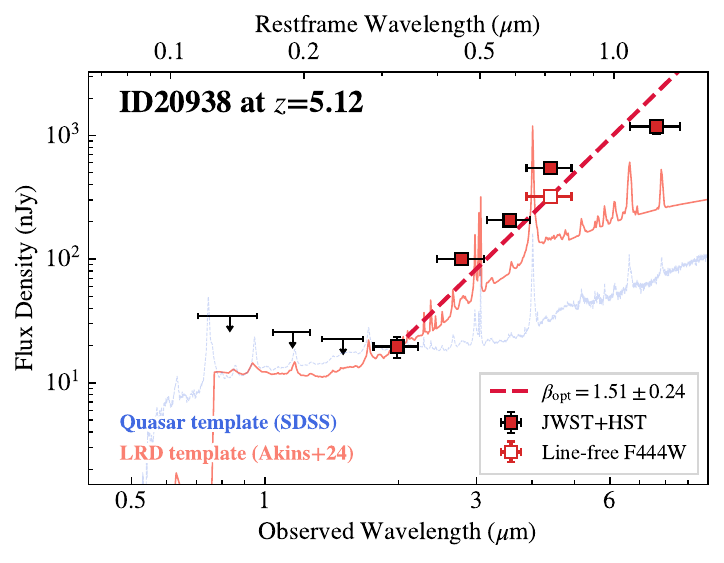}
    \caption{ SEDs of ID3878 and ID20938. Both sources are detected in the MIRI F770W band. ID3878 also has MIRI F1800W coverage but shows no detection, so we place a $3\sigma$ upper limit. The measured optical slopes are shown as red dashed lines. The LRD template from \cite{Akins2024} is shown for reference (red lines), and the quasar template from \cite{VandenBerk2001} is provided for comparison (blue dashed lines).  }
    \label{fig:lrd_like_SED_MIRI}
\end{figure*}

\subsubsection{UV magnitude and optical continuum slopes}

\begin{figure*}
    \includegraphics[width=0.33\textwidth]{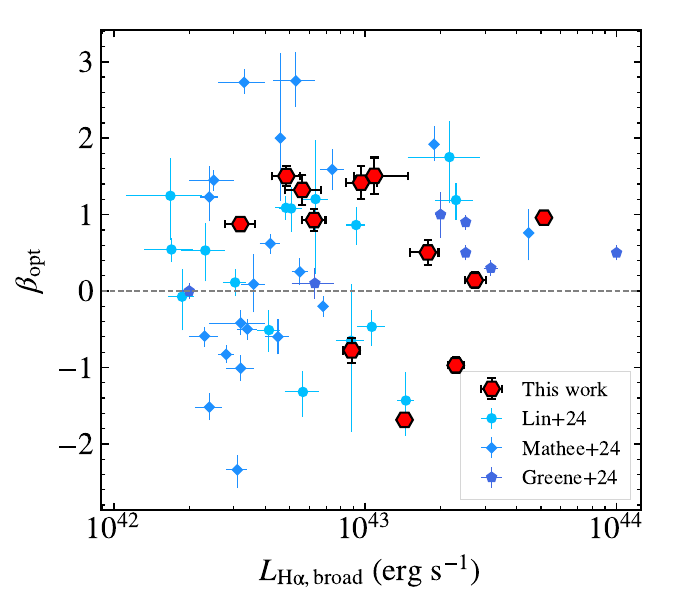}
    \includegraphics[width=0.33\textwidth]{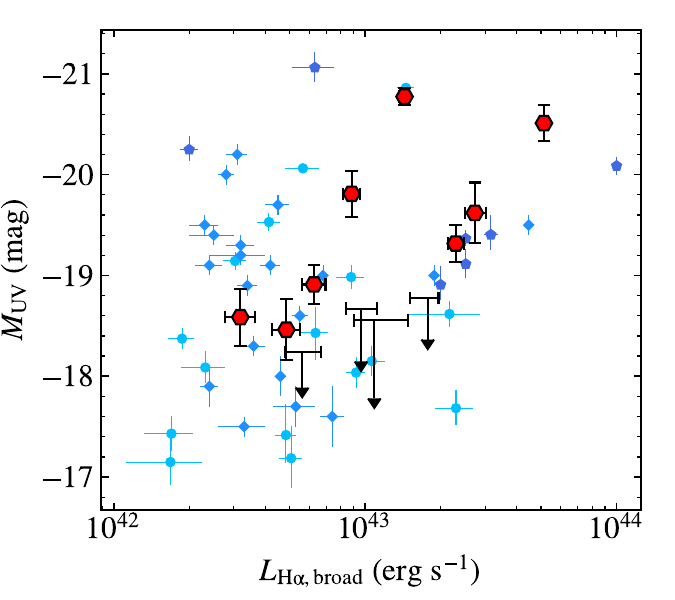}
    \includegraphics[width=0.33\textwidth]{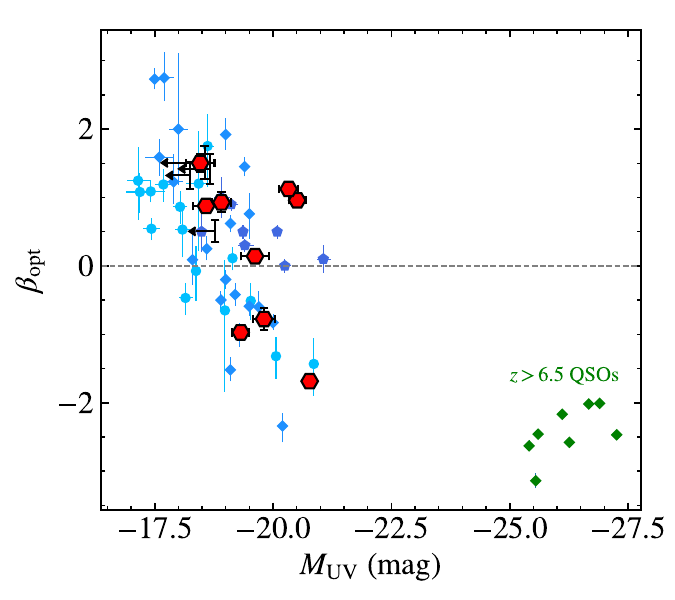}
    \caption{Broad H$\alpha$ luminosity $L_{\rm H\alpha, broad}$ vs. optical continuum slope $\beta_{\rm opt}$, $\beta_{\rm opt}$ vs. $M_{\rm UV}$, and $M_{\rm UV}$ vs. $L_{\rm H\alpha, broad}$ for broad-line AGNs. In the \textit{left} and \textit{middle} panels, when considering $L_{\rm H\alpha, broad}$, we exclude the $z=7.646$ broad H$\beta$ emitter in this work, as well as the $z=8.5$ broad H$\beta$ AGN of \cite{Greene2024}. In the \textit{right} panel, both broad H$\beta$ emitters are included. We also include $M_{1450}$ and $\beta_{\rm opt}$ for $z>6.5$ quasars presented by \cite{Yang2023}.
    }
    \label{fig:beta_Lha_Muv}
\end{figure*}

We estimate the observed rest-frame UV magnitude ($M_{\rm UV}$) based on the  
F115W magnitude, which corresponds to the rest-frame $1640-1900$ \AA\ for objects at $z = 5-6$. For \hbagn, we use the F150W magnitude to estimate the observed $M_{\rm UV}$, corresponding to a rest-frame wavelength of $\sim$1728 \AA. 

To measure the optical continuum slope (\betaopt), we first estimate the expected line-induced F444W photometry using the best-fit Gaussian profiles, as described in \S\ref{sec:line_profile}. We then subtract this from the original F444W photometry to obtain the line-free F444W photometry. We fit the power law from rest-frame 3000\AA\ to the longest wavelength, using JWST photometry only. Note that this measured \betaopt\ accounts for potential Balmer breaks.  We exclude the band containing [\ion{O}{3}] and use the line-free F444W photometry, ensuring that the photometry used is purely from the continuum.  The \betaopt\  values are listed in Table~\ref{tab:property}. 

We find that the broad-line AGNs in this work span a wide range of optical continuum slopes, from $\beta_{\rm opt} = -1.04$ to $1.50$. Of the 13 broad-line AGNs, 10 have $\beta_{\rm opt} > 0$ (77\%). The remaining three have $\beta_{\rm opt} < 0$ (23\%). We show two examples of $\beta_{\rm opt} < 0$ sources in Figure \ref{fig:quasar_like_SED} and two examples of the $\beta_{\rm opt} > 0$ sources in Figure \ref{fig:lrd_like_SED_MIRI}. The two sources in Figure \ref{fig:lrd_like_SED_MIRI} also have MIRI coverage, which will be discussed in the following section.  We compare the SEDs of these sources with the LRD template from \cite{Akins2024} and the quasar template from \cite{VandenBerk2001}.  We find that the SEDs of the sources with $\beta_{\rm opt} > 0$ resemble the LRD template, while the sources with $\beta_{\rm opt} < 0$ are more similar to the quasar template. We caution that $\beta_{\rm opt} = 0$ is an empirical and simplified yet necessary classification of these high-$z$ AGNs. Among the three $\beta_{\rm opt} < 0$ objects, ID27947 shows a strong \ion{He}{1} $\lambda$7067 line with broad components (right panel of Figure \ref{fig:GrismHa_withContinuum}).  Its $M_{\rm UV}$ is about --20.77 mag, placing it in the faintest regime of low-luminosity quasars at $z \sim 6$ \cite[$M_{1450}\approx-20.91$ at $z\approx 6.13$, ][]{Matsuoka2019}. The SEDs of all the AGN samples in this paper are presented in Appendix \S\ref{sec:sed_all}.

Our analysis reveals no significant correlation between $\beta_{\rm opt}$, $L_{\rm H\alpha, broad}$, and $M_{\rm UV}$, as shown in Figure \ref{fig:beta_Lha_Muv}. In the right panel of Figure \ref{fig:beta_Lha_Muv}, we compare the $M_{\rm UV}$ and $\beta_{\rm opt}$ of low-luminosity AGNs with those of quasars at $z>6.5$ \citep{Yang2023}. The $\beta_{\rm opt}$ values for quasars are derived by fitting their optical continua with a combination of a power-law model and an empirical optical \ion{Fe}{2} template. The power-law slopes may be degenerate with the \ion{Fe}{2} template. Nonetheless, these UV-luminous quasars exhibit very blue optical slopes, with an average $\beta_{\rm opt} \approx -2.4$. Broad-line AGNs with $\beta_{\rm opt} < 0$ are primarily found in the relatively UV-bright regime ($M_{\rm UV} < -19$), but their $\beta_{\rm opt}$ values are not as blue as those of quasars. The number of optically red yet UV-bright sources (e.g., \hbagn) is small.

\subsubsection{MIRI photometry}\label{sec:miri}

Of the 13 broad-line AGNs in our sample, 2 have MIRI coverage. Both ID3878 and ID20938 are detected in F770W, and ID3878 is also covered by F1800W. We list their MIRI photometry in Table \ref{tab:miri}.

As shown in Figure \ref{fig:lrd_like_SED_MIRI}, the F770W fluxes of both objects, corresponding to rest-frame $\sim 1$\,\micron, are slightly lower than the power-law extrapolation from the optical continua. The $3\sigma$ upper limits on F1800W for ID3878 provide a strong constraint on its IR color, showing that the IR SED does not continue to rise. It is consistent with the rest-frame IR colors of $z\approx3-4$ LRDs reported in the literature \citep{Wang2024b, Juodzbalis2024, Labbe2024}. This is in good agreement with the hypothesis that LRDs lack hot dust or have a lower hot dust fraction compared to typical quasars \citep{Casey2024, Leung2024}. 

\begin{table}[htbp]
    \centering
    \begin{tabular}{ccc}
    \hline
       ID  & $f_{\rm F770W}$  & $f_{\rm F1800W}$\\
        & (nJy) & (nJy) \\
    \hline
        3878 &  $750\pm83$ & $<1311$ \\
        20938 &  $1184\pm152$  & -- \\   
    \hline
    \end{tabular}
    \caption{MIRI photometry of the two broad-line AGNs, as presented in Section~\ref{sec:miri}. For ID3878, a $3\sigma$ upper limit is reported for $f_{\rm F1800W}$.}
    \label{tab:miri}
\end{table}

\subsection{Variability}

The COSMOS-3D program repeats NIRCam/F115W observations in regions where existing COSMOS-Web images are available. The F115W images of COSMOS-Web in this footprint were taken between January 2023 and January 2024, while the F115W images of COSMOS-3D were obtained between November 2024 and December 2024.  For the broad-line AGN sample in this paper, F115W corresponds to rest-frame $1700-1900$\AA\ for objects at $z=5-6$ and $\sim1300$\AA\ for ID13852 at $z=7.646$. The observed-frame time interval between the two epochs ranges from 350 to 715 days, which corresponds to a rest-frame time interval of 50 to 82 days.

We investigate the rest-frame UV variability of the sample by comparing their $r = 0.15\arcsec$ aperture photometry at the two epochs. The variability measurements are cross-checked using the difference images (see Appendix \ref{sec:variability_appendix} for further details).  Among the 13 AGNs, 12 have flux and magnitude variations from the two methods consistent with zero within 1$\sigma$, and ID27974 shows variations consistent with zero within 2$\sigma$. It suggests that these objects exhibit no significant variability over $\sim60$ rest-frame days, consistent with the findings by \cite{Kokubo2024, Tee2024}.

\subsection{Luminosity function of broad-line \ha\ emitters}\label{sec:ha_lf}

We compute the total H$\alpha$ LF for the 12 broad-line H$\alpha$ emitters, incorporating both the broad and narrow components. We also compute the LF for their broad H$\alpha$ component.  We follow the methodology outlined by \cite{Lin2024}.

We calculate the \ha\ LFs using the direct 1/$V_{\rm max}$ method \citep{Schmidt1968}. As a complement, we also calculate the LF using Lynden-Bell's $C^-$ method \citep{Lynden-Bell1971}. Lynden-Bell's $C^-$ method is an unbinned maximum likelihood estimator of the cumulative luminosity function for flux-limited truncated data. We present LF measurements using $C^-$ method in Appendix \ref{sec:appendix_lf_cm}.  For the $1/V_{\rm max}$ method, the LF is computed as

\begin{equation}
\Phi(L)=\frac{1}{d \log L} \sum_i \frac{1}{C_i V_{\max , i}}.
\end{equation}
For the total H$\alpha$ LF, $L$ is the total H$\alpha$ luminosity of each LF bin, $C_i$ accounts for the completeness correction, and $V_{\max,i}$ is the maximum survey volume for the $i$-th broad-line H$\alpha$ emitter. For the broad H$\alpha$ LF, $L$ refers to the luminosity of the broad emission lines. The details about $V_{\max,i}$ and completeness correction are presented in Appendix \ref{appendix:lf}.  We conduct Monte Carlo experiments to estimate the uncertainties in the LFs. These experiments incorporate Poisson noise for small number statistics \citep{Gehrels1986} and propagated uncertainties in the luminosity measurements.

\begin{figure}[!t]
    \centering
    \includegraphics[width=1\linewidth]{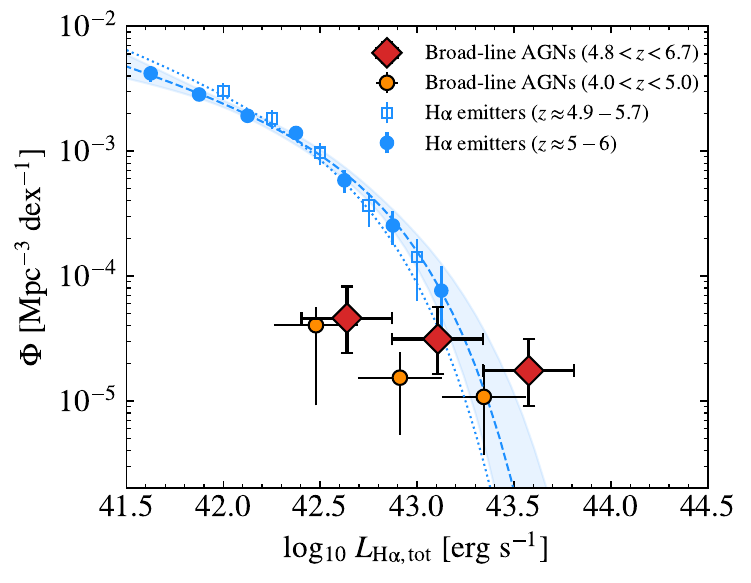}
    \caption{H$\alpha$ (broad + narrow) LF of broad-line AGNs at $4.8<z<6.7$. We also present the H$\alpha$ LF of broad-line AGNs at $4.0<z<5.0$ in \cite{Lin2024}. The H$\alpha$ LFs of star-forming galaxies from \citet{CoveloPaz2024} and \citet{Lin2025}, which exclude AGNs, are shown as open blue squares and filled blue dots, respectively. The blue dashed lines and shaded regions represent the best-fit model from \cite{Lin2025}, while the dotted lines indicate the best-fit model from \cite{CoveloPaz2024}.}
    \label{fig:LF_tot}
\end{figure}

The measured total H$\alpha$ LF and broad-line H$\alpha$ LF are presented in Table \ref{tab:LF}. We compare the total H$\alpha$ LF of broad-line AGNs with that of star-forming galaxies at $z\approx5-6$ from \cite{CoveloPaz2024} and \cite{Lin2025} in Figure \ref{fig:LF_tot}. At $L_{\rm H\alpha, tot} \approx 10^{43.2}$\,\si{erg\,s^{-1}}, the number density of broad-line AGNs is comparable to that of star-forming galaxies. However, at $L_{\rm H\alpha, tot} \approx 10^{43.6}$\,\si{erg\,s^{-1}}, the number density of star-forming galaxies drops significantly, such that the H$\alpha$ emitters at these luminosities are dominated by broad-line AGNs. We note that the bright-end H$\alpha$ LFs of star-forming galaxies ($L_{\rm H\alpha} > 10^{43.5}$\,\si{erg\,s^{-1}}) are based on extrapolation from the best-fit Schechter function, and the full COSMOS-3D dataset across 0.33\,\si{deg^{2}} will provide a more robust constraint on it.

\begin{table}[!t]
	\begin{center}
		\begin{tabular}{cccc}
        \hline
        \hline
			$\log L_{\rm H\alpha}$ & $\Delta \log L_{\rm H\alpha}$ & $N$ & $\Phi (10^{-5} {\rm Mpc}^{-3} {\rm dex}^{-1})$ \\
			\hline
			$42.638$ & $0.469$ & 4 & $4.61^{+3.64}_{-2.20}$ \\
			$43.107$ & $0.469$ & 4 & $3.13^{+2.47}_{-1.50}$ \\
			$43.576$ & $0.469$ & 4 & $1.75^{+1.38}_{-0.84}$ \\
        \hline
        \hline
        $\log L_{\rm H\alpha, broad}$ & $\Delta \log L_{\rm H\alpha, broad}$ & $N$ & $\Phi (10^{-5} {\rm Mpc}^{-3} {\rm dex}^{-1})$ \\
		\hline
	$42.671$ & $0.436$ & 4 & $4.96^{+3.92}_{-3.26}$ \\
	$43.107$ & $0.436$ & 5 & $3.81^{+2.58}_{-2.22}$ \\
	$43.543$ & $0.436$ & 3 & $1.43^{+1.39}_{-0.78}$ \\
        \hline
		\end{tabular}
    \caption{The \ha\ (narrow + broad) LF and broad-line \ha\ LF as shown in Figure \ref{fig:LF_tot} and \ref{fig:LF_broad}.}
    \label{tab:LF}
	\end{center}
\end{table}

We compare the total H$\alpha$ LF in this work to the total H$\alpha$ LF of broad-line AGNs at $z \approx 4 - 5$ in \cite{Lin2024}, as shown in Figure \ref{fig:LF_tot}. The H$\alpha$ LF of broad-line AGNs in \cite{Lin2024} is constructed using the same selection method as in this work, based on 25 independent JWST fields with a total area of 275 arcmin$^{2}$, so the impact of cosmic variance is small. We find that the LF of broad-line AGNs at $z \approx 4.8 - 6.7$ is consistently higher than that at $z \approx 4 - 5$ across all luminosity bins, despite the large uncertainties caused by the small sample size. This trend may indicate a potential increase in the number density of broad-line AGNs from $z \approx 4 - 5$ to $z \approx 5 - 7$.  However, we also caution the cosmic variance in the limited volume of the current COSMOS-3D data. The cosmic variance in this work is more than twice that of the ASPIRE survey.

\begin{figure}[!t]
    \centering
    \includegraphics[width=1\linewidth]{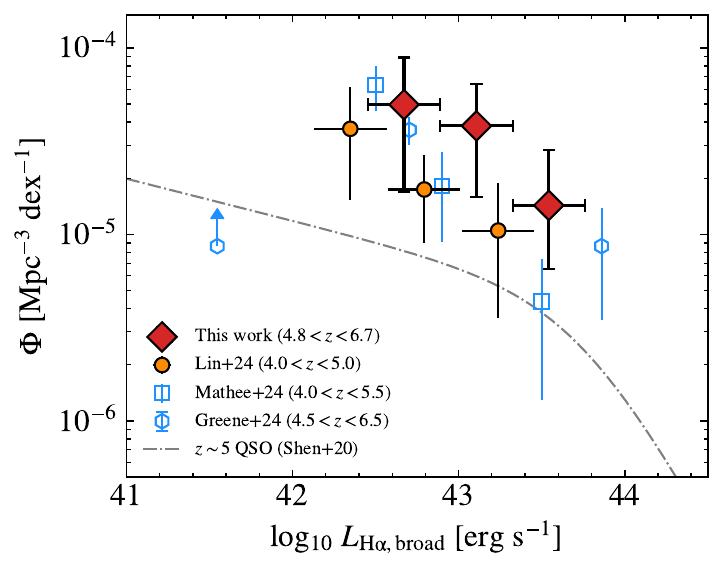}
    \caption{Broad-line \ha\ LF at $5<z<6.7$. We also present the broad-line \ha\ LFs from broad-line AGNs in \cite{Matthee2024, Greene2024, Lin2024}. The gray dotted-dashed line represents the extrapolation of the quasar bolometric LF at $z \sim 5$, assuming the relationship between quasar broad-line H$\alpha$ luminosity and bolometric luminosity. }
    \label{fig:LF_broad}
\end{figure}

The broad H$\alpha$ LF, in the context of broad H$\alpha$ LFs of AGNs from the literature, is shown in Figure \ref{fig:LF_broad}. We convert the broad H$\alpha$ luminosity for $z \sim 5$ quasar bolometric LFs from \cite{Shen2020}, assuming the relationship among the bolometric luminosity $L_{\rm bol}$, the rest-frame 5100\AA\ luminosity, and the AGN-induced H$\alpha$ luminosity from \cite{Greene2005}, with the bolometric correction from \cite{Richards2006}. We caution that the assumptions applied to type-1 AGNs in the local Universe may not directly apply to broad-line AGNs at high redshift. In this work, we use the conversion only for a qualitative comparison of number densities. We find that at $L_{\rm H\alpha, broad} < 10^{43.5}$\,\si{erg\,s^{-1}}, the number density of broad-line AGNs is about 6 times higher than the extrapolation of quasars. This result is consistent with the literature \citep{Matthee2024, Greene2024, Lin2024} and further confirms the prevalence of broad-line AGNs at $z>5$.

\subsection{The abundance of broad \hb\ emitters at  $z > 7$}

\begin{figure}
    \centering
    \includegraphics[width=1\linewidth,trim={0 18pt 0 0}]{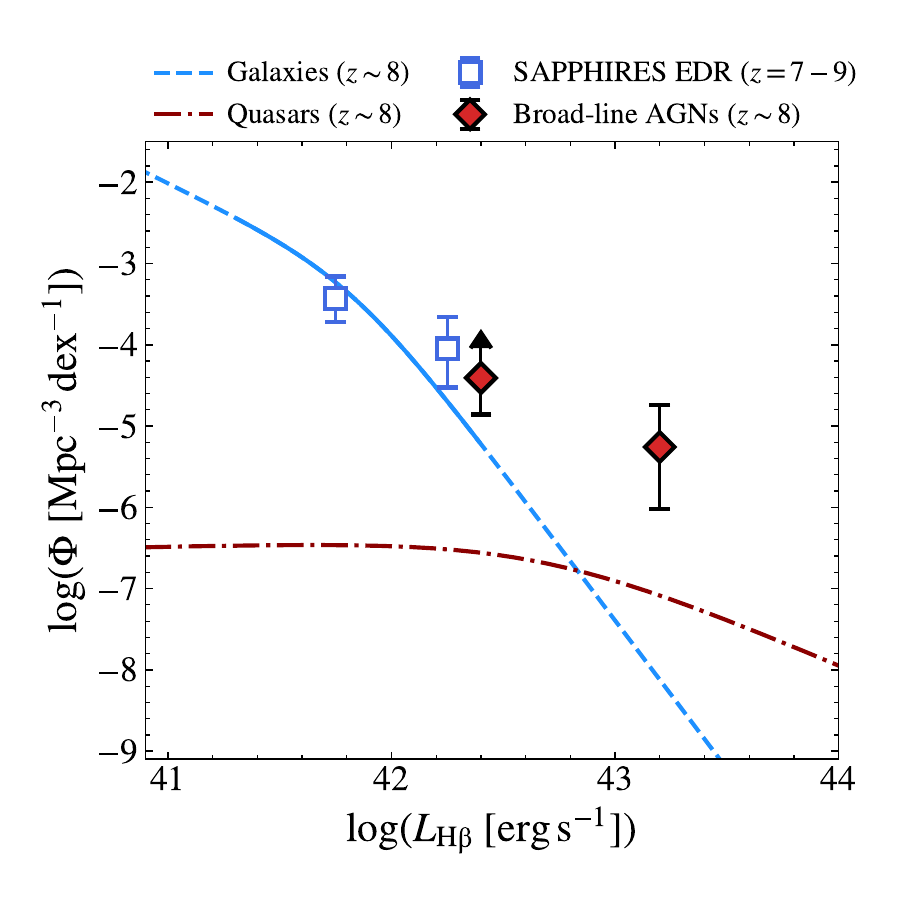}
    \caption{Total \hb\ luminosity (broad + narrow) function of broad-line AGN at $z\sim 8$ (red diamonds).
    This is derived based on \hbagn, the $z=8.48$ AGN in \cite{Fudamoto2025}, and the $z=8.50$ AGN in \cite{Kokorev2023}. The \hb\ LF scaled from [\ion{O}{3}] LF at $z=7.5-9$ \citep{Meyer2024} is shown as the blue line, where the dashed parts are extrapolations from the best-fit [\ion{O}{3}] LF in double power law. The measured \hb\ LF at $z\sim8$ based on the \hb+[\ion{O}{3}] emitters from SAPPHIRES EDR \citep{Sun2025} is shown by the blue squares. 
    The red dashed-dotted line is the quasar \hb\ LF prediction at $z\sim8$ based on the TRINITY empirical models \citep{ZhangH2024}.
}
    \label{fig:LF_Hb}
\end{figure}

In the 127 arcmin$^{2}$ covered by the first 10\% of the COSMOS-3D data, we detect a bright broad H$\beta$ emitter at $z=7.647$ (\hbagn). This object has a total H$\beta$ luminosity of $\log\left(L_{\rm H\beta, tot}/\,{\rm erg\, s^{-1}}\right) \approx 43.33\pm0.09$, and a narrow H$\beta$ luminosity of $\log\left(L_{\rm H\beta. narrow}/\,{\rm erg\, s^{-1}}\right) \approx 42.55\pm0.13$.  Its narrow H$\beta$ luminosity also exceeds that of the most H$\beta$-luminous galaxies found in the total 124 arcmin$^2$ FRESCO fields \citep[$\log\left(L_{\rm H\beta}/\,{\rm erg\, s^{-1}}\right) \approx 42.22\pm0.05$,][]{Meyer2024}.

We derive the total H$\beta$ LF of broad-line AGNs at $z\sim8$ by combining \hbagn, the  $z=8.48$ broad-line AGN in \cite{Fudamoto2025}, and the $z=8.50$ broad-line AGN in \cite{Kokorev2023}.  For reference, we also include two \hb\ LFs of galaxies at similar redshifts in Figure \ref{fig:LF_Hb}.  One is scaled from the [\ion{O}{3}] LF at $z=7.5-9$ in \cite{Meyer2024}, and the other is measured based on grism-selected \hb+[\ion{O}{3}] emitters in JWST Cycle 3 SAPPHIRES Early Data Release \citep[EDR;][]{Sun2025}. The measurements of AGN and galaxy \hb\ LFs are detailed in Appendix \ref{appendix:hb_lf}.

As shown in Figure \ref{fig:LF_Hb}, \hb\ emitters are dominated by AGNs at $L_{\rm H\beta} \geq 10^{42.4}\,{\rm erg\,s^{-1}}$. 
In this luminosity regime, the number density of broad-line AGN greatly exceeds the extrapolation from the galaxy LF.
The observed number density of broad-line AGNs at $z\sim8$ is also about 100 times higher than the predicted \hb\ LF of quasars at $z\sim8$ from the TRINITY empirical model \citep[][see details in Appendix~\ref{appendix:hb_lf}]{ZhangH2024}.

The estimated number density derived from \hbagn\ is $(6.05^{+13.93}_{-5.00}) \times 10^{-6} \, \text{Mpc}^{-3}$.  
Extrapolating the $z \sim 7$ UV LF \citep{Matsuoka2023}, we estimate the number density of faint quasars with $M_{1450} > -23$ at $z \approx 7.5$ to be $2 \times 10^{-8} \, \text{Mpc}^{-3}$. This value is 2 orders of magnitude lower than that derived from \hbagn. If this holds, there would be only a 1\% chance of finding faint quasars within the current COSMOS-3D survey volume. 
The discovery of \hbagn\ suggests that UV-faint type-1 AGNs at $z > 7$ are much more abundant than predicted by the quasar LF. These AGNs may be reddened and were missed by previous UV selection methods. They may not belong to the same population as the UV-selected faint quasars identified before JWST.  

We note that within the 124 arcmin$^2$ FRESCO survey, there is also a broad \hb\ emitter, GNz7q \citep{Fujimoto2022}, with an \hb\ luminosity of $\log\left(L_{\rm H\beta, tot}/{\rm erg\,s^{-1}}\right) = 41.96 \pm 0.08$ \citep{Meyer2024}. Although GNz7q has a different SED shape compared to the so-called LRDs, we emphasize that the broad-line selection is independent of the AGN's SED shape. Combining the FRESCO survey and the first 10\% of the COSMOS-3D survey (and also the SAPPHIRES EDR), we empirically infer that there is likely at least 1 broad \hb\ emitter at $z>7$ over approximately 120 arcmin$^2$. The full COSMOS-3D dataset will provide crucial insights into the bright-end \hb\ LF and the abundance of AGNs at $z>7$.

\section{Discussion}\label{sec:discussion}

\subsection{Implications of the Balmer absorption}

Two sources (ID3878 and ID17455) in our sample exhibit \ha\ absorption, comprising 15\% of the total. This fraction is consistent with the value reported by \cite{Lin2024}. The two, along with the three in \cite{Lin2024}, all have $\beta_{\rm opt} > 0$ ($0.5-1.5$).   In both studies, the AGN samples are selected through a broad-line search without any photometric selection, resulting in a sample that includes both optically blue and red sources. Therefore, we conclude that Balmer absorption is observed exclusively in optically red broad-line AGNs with $\beta_{\rm opt} > 0$. This association aligns with the hypothesis that Balmer absorption originates from dense gas surrounding the BLRs. The high-density gas would also cause Balmer breaks and result in visually red $\beta_{\rm opt}$ \citep{Inayoshi2024, Ji2025, Naidu2025, deGraaff2025}. The MIRI photometry of ID3878 reveals a decreasing rest-frame 1-2 \,\micron\ continuum in the $f_\lambda$ space (or flat in $f_\nu$ space), similar to those of sources in \cite{Naidu2025, deGraaff2025}. This continuum shape can be generally explained by an intrinsic AGN spectrum attenuated by high-density hydrogen.

The current photometric data in the COSMOS field are insufficient to confirm the presence of a Balmer break. Future NIRSpec Prism observations will be crucial for confirming the presence of Balmer breaks and validating the models.

\subsection{Implications for rest-frame UV Variability}

For AGNs, the characteristic timescale and amplitude for variability can be derived from the empirical relation as a function of bolometric luminosity ($L_{\rm bol}$), \MBH\ and the rest-frame wavelengths \citep{MacLeod2010, Burke2023}. The variability is usually analyzed using a structure function (SF) approach \citep{Hughes1992, Collier2001, Bauer2009, Kozlowski2010},
where the SF represents the RMS magnitude difference as a function of the time lag ($\Delta t$) between measurements. The long-timescale variability is characterized by the asymptotic variability amplitude (${ \rm SF}_\infty$) and the characteristic timescale ($\tau$).

We estimate the predicted UV variability using the parameterization from \cite{Burke2023}. For AGNs in our sample (\MBH$\approx 10^{7.0}-10^{8.6}\ M_\odot$, Table \ref{tab:property}), assuming Eddington ratios of 0.1, the expected variability amplitude is ${\rm SF}_\infty \approx 0.11 -0.18$ mag at rest-frame 1700\,\AA, with  $\tau \approx 134-221$ days. For timescales of $0.1<\Delta t/\tau<1$, the ${\rm SF}(\Delta t)$ scales with a power law with an index of 0.5 based on the damped random walk (DRW) model \citep[e.g.,][]{MacLeod2010, MacLeod2012, Kozlowski2016}. If adopting this prescription, a variation over $\Delta t= 60$ days corresponds to about  ${\rm SF}(\Delta t)=0.06-0.11$ mag. 

If the UV light of the broad-line emitters in this work is dominated by AGN emission and the process follows the DRW model, the expected magnitude variation $\Delta m$ between the two epochs should be a value randomly drawn from a Gaussian distribution with a standard deviation of ${\rm SF}(\Delta t)$, while also incorporating the photometric uncertainties and scatters for nonvariable sources. However, the photometric uncertainties listed in Table \ref{tab:var_f115w} typically range from 0.2 to 0.9 mag. It suggests that the F115W images are not deep enough in most regions to detect variability of ${\rm SF}(\Delta t = 60 \, {\rm days}) \approx 0.1$ mag. 

One may caution about the impact of \MBH\ in the prediction above, as \MBH\ estimates for high-redshift AGNs have been highly debated in the literature.\citep[e.g.,][]{Kokubo2024a, Juodzbalis2024, Naidu2025, Rusakov2025, Juodzbalis2025}. For instance, dense gas surrounding the BLRs may broaden the Balmer lines, as proposed in \cite{Rusakov2025} and \cite{Naidu2025},  leading to an overestimation of \MBH\ by $1-2$ dex when derived directly from the observed FWHMs. In such cases, the typical \MBH\ of the sample in this work could be as low as $10^6 M_\odot$. An AGN with \MBH$=10^6 M_\odot$ and an Eddington ratio of 0.1 is expected to show a variability amplitude of ${\rm SF}(\Delta t = 60 \, {\rm days}) = 0.27$ mag.  If such an AGN is accreting with an Eddington ratio of 1, the variability amplitude would decrease to ${\rm SF}(\Delta t = 60 \, {\rm days}) = 0.13$ mag.  In most regions of the shallow F115W images, such magnitude differences would also be overwhelmed by photometric uncertainties and therefore remain undetectable.

We thus conclude that the F115W images of COSMOS-3D and COSMOS-Web may not be deep enough to detect the variability of broad-line AGNs at $z>5$, even assuming the UV light is entirely dominated by AGN emission. The expected variability could be diluted if the UV light includes contributions from galaxies. The physical properties of the BHs (e.g., mass, accretion rate, etc.) would further complicate the interpretation. On the other hand, this suggests that any detection of variability in $z > 5$ broad-line AGNs with COSMOS-3D would provide novel insights into the BH activities in the early Universe. It would serve as an independent approach to testing the existing models.

\section{Conclusion}

In this work, we present a study of 13 broad-line AGNs at $z>5$ selected from the first observations of the COSMOS-3D grism survey.  The sample includes 12 broad \ha\ emitters at $z\approx 5-6$ and one broad \hb\ emitter at $z=7.646$. Among them, four objects exhibit F444W magnitudes brighter than 24 mag.  The $z\approx 7.646$ broad \hb\ emitter, \hbagn, has an F444W magnitude of 23.64, making it among the brightest broad-line AGNs discovered by JWST at $z>7$. We analyze the SED shape, spectral profiles, variability, and LF of the broad-line AGNs. Our main conclusions are as follows.

\begin{itemize}
    \item We find two AGNs with reddened optical continua detected in the grism spectra, including \hbagn. One AGN (ID27974) at $z=5.03$ exhibits a broad \ion{He}{1} 7067 emission line and blue optical continuum (\betaopt$<0$). The source has $M_{\rm UV}\approx-20.77$ mag, within the faintest regime of low-luminosity quasars. Additionally, we find that two AGNs show blue-shifted \ha\ absorption.

    \item Among the 13 AGNs, 10 have \betaopt$>0$ and 3 have \betaopt$<0$. The SEDs of \betaopt$<0$ resemble the SEDs of quasars, while the sources with \betaopt$>0$ are more similar to LRDs.   The \betaopt\, broad-line luminosity $L_{\rm H\alpha, broad}$ and $M_{\rm UV}$ are not clearly correlated. The \betaopt$<0$ sources are primarily UV bright, but their \betaopt\ are not as blue as UV-luminous quasars.

    \item  Two of the AGNs have MIRI coverage, and one is observed in both F770W and F1800W. Although it is undetected in F1800W, this nondetection imposes a strong upper limit on its near-IR color, indicating that its SED is unlikely to rise steeply from the optical to the near-IR.

    \item We examine the rest-frame UV variability by comparing the COSMOS-3D and COSMOS-Web F115W images. The time intervals span from 350 to 715 days, corresponding to 50-82 days in the rest-frame. None of the AGNs show significant variability. The flux variation between the two epochs is consistent with zero within 1$\sigma$-2$\sigma$. Based on empirical models of AGN variability, we find that the depth of the F115W images is not sufficient to detect the expected variability of broad-line AGNs at $z>5$, even assuming the UV light is entirely AGN-dominated.

    \item We compute the total (broad + narrow) \ha\ LFs for the 12 AGNs at $z=5-6$. We find that the number density of AGNs at $z>5$ is higher than that at $z=4-5$, indicative of potential redshift evolution in AGN abundance.  The broad \ha\ LF is 7 times higher than $z\sim5$ quasar LF extrapolation. 

    \item We derive the \hb\ LF at $z \sim 8$ by combining \hbagn\ with two $z \sim 8$ broad-line AGNs from the literature. Compared to the LF of galaxies, broad-line AGNs dominate at the bright end ($L_{\rm H\beta} > 10^{42.4}$\,\si{erg\,s^{-1}}). The number density of broad-line AGNs at $z \sim 8$ indicated by \hbagn\ is 2 orders of magnitude higher than the value extrapolated from the quasar LF.

    \item Combining our sample with the broad-line selected AGNs from \cite{Lin2024}, we conclude that Balmer absorbers are exclusively observed in AGNs with \betaopt$ > 0$. This finding is consistent with the hypothesis that Balmer absorption originates from dense gas surrounding the BLR and may be associated with Balmer breaks. Future NIRSpec follow-up observations will further investigate the presence of Balmer breaks.

\end{itemize}

The paper demonstrates the capability of the COSMOS-3D dataset in studying the nature of broad-line AGNs at $z>5$. The full COSMOS-3D grism survey will cover a total area of 0.33 deg$^2$, providing a statistically significant large sample. Combined with the existing COSMOS-Web data and 482 arcmin$^2$ of COSMOS-3D MIRI parallel observations with F1000W and F2000W, it will offer new insights into AGN photometric properties, the occurrence of Balmer absorbers, UV variability, and the AGN abundance at $z=5-8$.
 
\section*{Acknowledgments}

We thank the anonymous referee for the constructive comments. We thank Stacey Alberts and Andras Gaspar for constructing and providing the MIRI empirical PSFs. We also appreciate Yang Sun for her assistance with these. X.L. and X.F. acknowledge support from the NSF award AST-2308258. F.W. acknowledges support from  NSF award AST-2513040. X.L. and Z.C. acknowledge support from the National Key R\&D Program of China (grant no. 2023YFA1605600) and Tsinghua University Initiative Scientific Research Program (No. 20223080023). M.V.\ gratefully acknowledges financial support from the Independent Research Fund Denmark via grant numbers DFF 8021-00130 and  3103-00146 and from the Carlsberg Foundation via grant CF23-0417. A.K.I.\ was supported by JSPS KAKENHI Grant Number JP23H00131. T.S.T. is supported by Forefront Physics and Mathematics Program to Drive Transformation (FoPM), a World-leading Innovative Graduate Study (WINGS) Program, the University of Tokyo. S.E.I.B.~is supported by the Deutsche Forschungsgemeinschaft (DFG) under Emmy Noether grant number BO 5771/1-1. J.S.K. and S.H.~are supported for this work by NASA through grant JWST-GO-01727 awarded by the Space Telescope Science Institute, which is operated by the Association of Universities for Research in Astronomy, Inc., under NASA contract NAS 5-26555. R.D.~acknowledges support from the INAF GO 2022 grant ``The birth of the giants: JWST sheds light on the build-up of quasars at cosmic dawn'' and by the PRIN MUR ``2022935STW'', RFF M4.C2.1.1, CUP J53D23001570006 and C53D23000950006. JTS is supported by the Deutsche Forschungsgemeinschaft (DFG, German Research Foundation) - Project number 518006966. KK acknowledges support from DAWN fellowship and VILLUM FONDEN (71574). The Cosmic Dawn Center (DAWN) is funded by the Danish National Research Foundation under grant No. 140. The French contingent of the COSMOS team is partly supported by the Centre National d’Etudes Spatiales (CNES). OI acknowledge the funding of the French Agence Nationale de la Recherche for the project iMAGE (grant ANR-22-CE31-0007).

This work is based on observations made with the NASA/ESA Hubble Space Telescope and NASA/ESA/CSA James Webb Space Telescope. The data were obtained from the Mikulski Archive for Space Telescopes (MAST) at the Space Telescope Science Institute, which is operated by the Association of Universities for Research in Astronomy, Inc., under NASA contract NAS 5-03127 for JWST. These observations are associated with program \#1727 (COSMOS-Web),  \#1873 (PRIMER), \#5893 (COSMOS-3D) and \#6434 (SAPPHIRES). Support for program \#5893 and \#6434 was provided by NASA through a grant from the Space Telescope Science Institute, which is operated by the Association of Universities for Research in Astronomy, Inc., under NASA contract NAS 5-03127. The MIRI products presented herein were retrieved from the Dawn JWST Archive (DJA). DJA is an initiative of the Cosmic Dawn Center (DAWN), which is funded by the Danish National Research Foundation under grant DNRF140.

\section*{Data Availability}
The JWST data presented in this article were obtained from the Mikulski Archive for Space Telescopes (MAST) at the Space Telescope Science Institute. The specific observations analyzed can be accessed via \dataset[DOI: 10.17909/ys3r-yp43]{https://doi.org/10.17909/ys3r-yp43}. The MIRI data products can be accessed through \dataset[the DAWN JWST Archive]{https://dawn-cph.github.io/dja}.

\appendix
\counterwithin{figure}{section}
\counterwithin{table}{section}

\section{SEDs of the full sample}\label{sec:sed_all}

\begin{figure}
    \centering
    \includegraphics[width=0.325\textwidth]{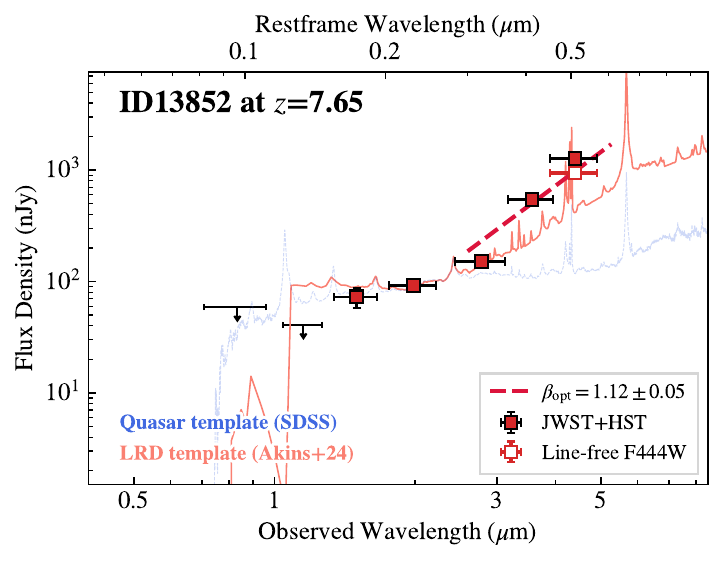}
    \includegraphics[width=0.325\textwidth]{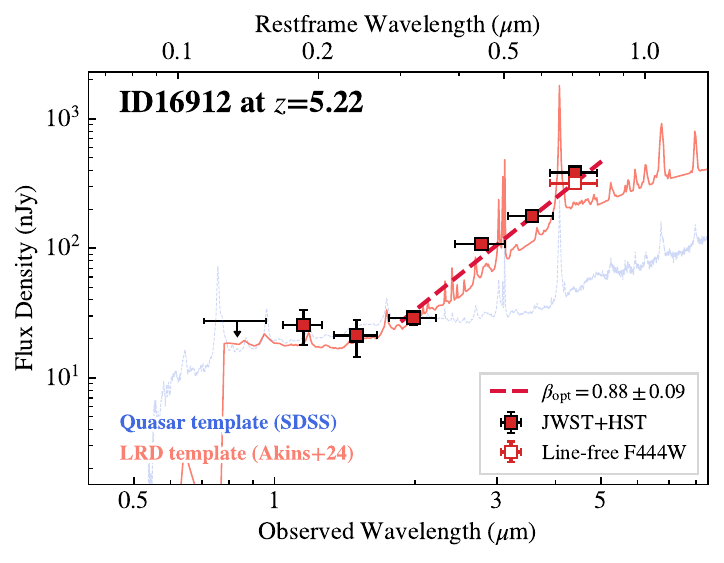}
    \includegraphics[width=0.325\textwidth]{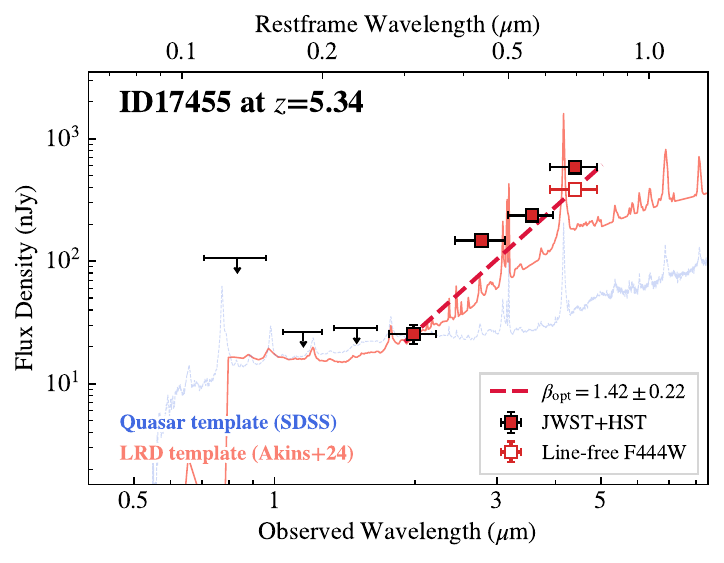}
    \includegraphics[width=0.325\textwidth]{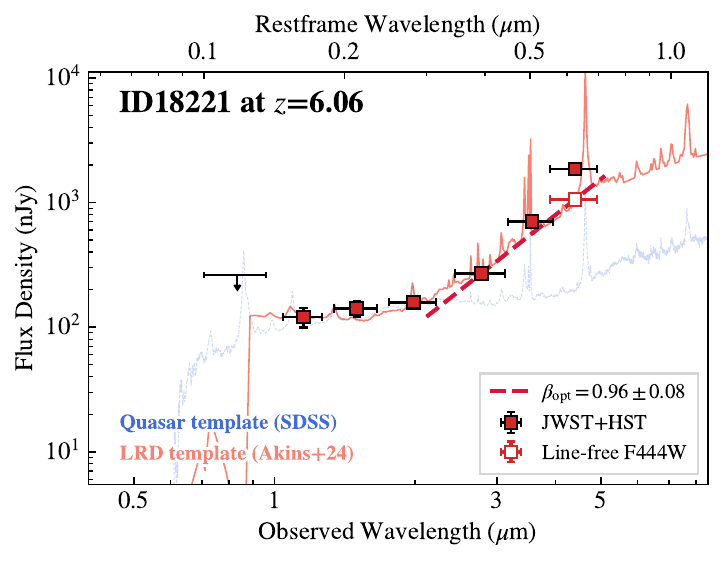}
    \includegraphics[width=0.325\textwidth]{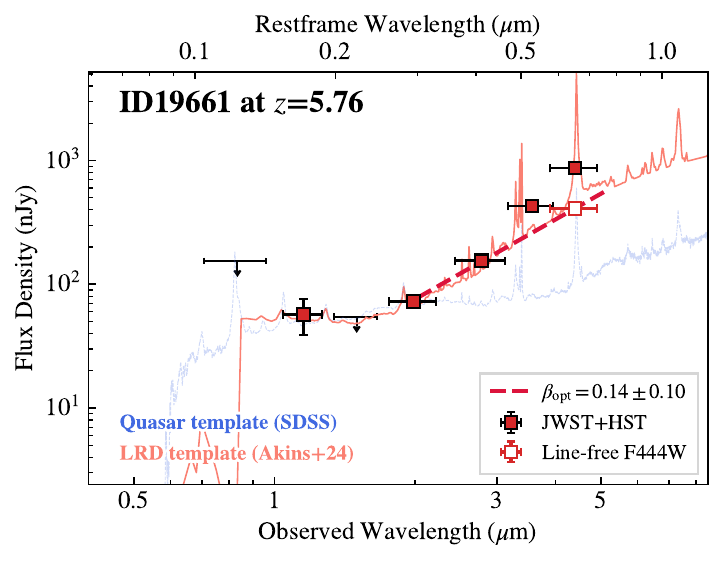}
    \includegraphics[width=0.325\textwidth]{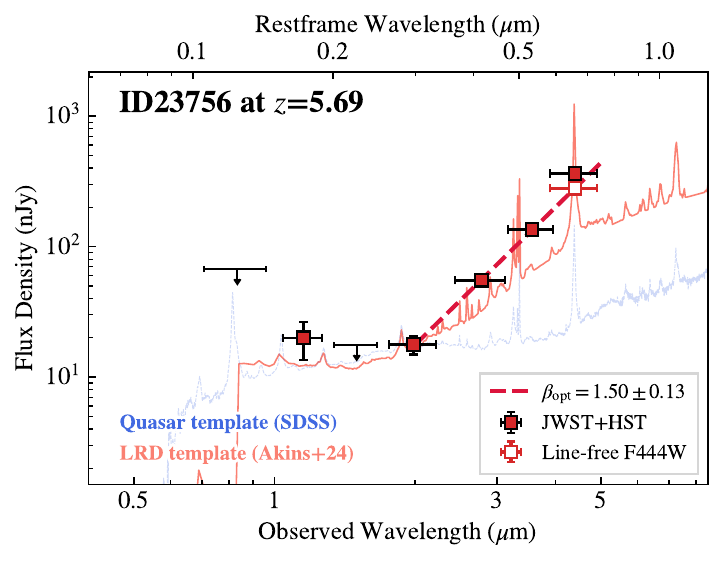}
    \includegraphics[width=0.325\textwidth]{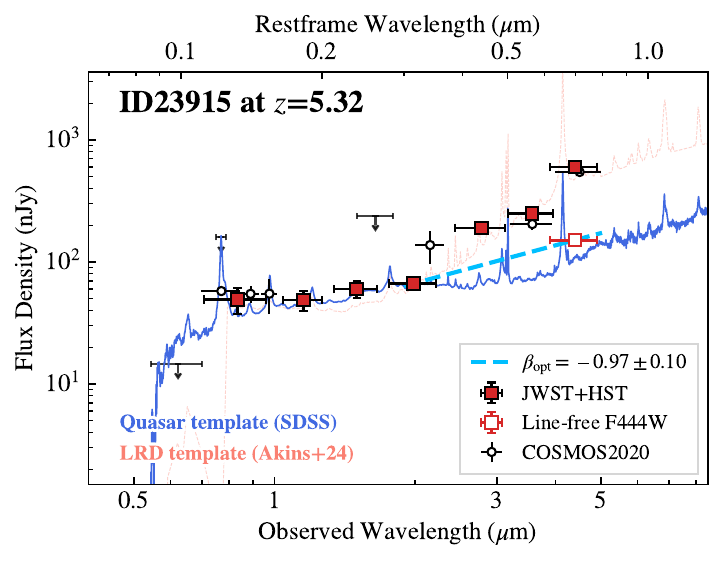}
    \includegraphics[width=0.325\textwidth]{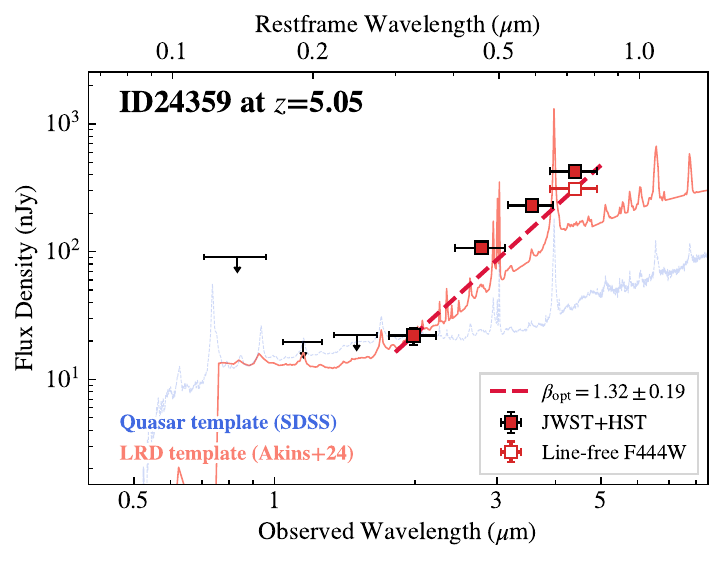}
    \includegraphics[width=0.325\textwidth]{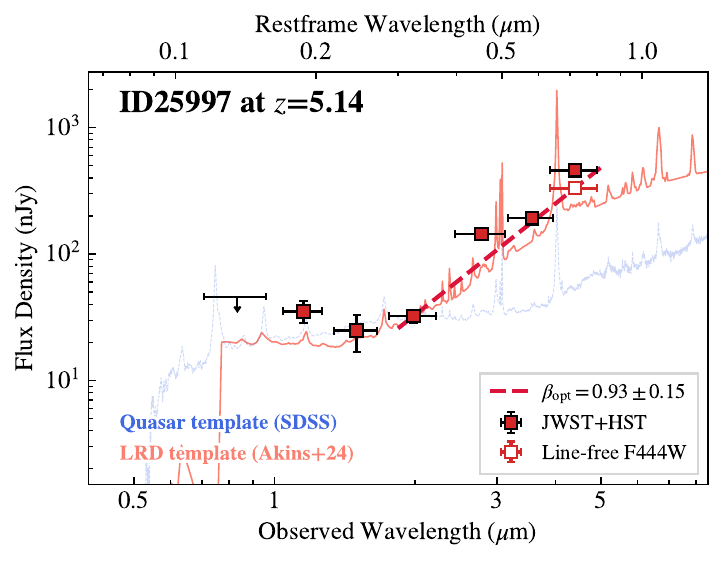}
    \caption{Similar to Figure \ref{fig:quasar_like_SED} and \ref{fig:lrd_like_SED_MIRI}, but show the SEDs of the remaining nine broad-line AGNs. For objects with $\beta_{\rm opt} > 0$, the red dashed line represents the best-fit optical continuum, with the \cite{Akins2024} LRD template shown as a red solid line for reference and the \cite{VandenBerk2001} quasar template as a blue dashed line for comparison. For objects with $\beta_{\rm opt} < 0$, we indicate the optical continuum using the blue dashed line, and highlight the \cite{VandenBerk2001} quasar template for reference. For ID19661, whose $\beta_{\rm opt}$ is consistent with zero within $1\sigma$, both the \cite{Akins2024} LRD template and the \cite{VandenBerk2001} quasar template are shown as solid lines.}
    \label{fig:sed_all}
\end{figure}
We present the multiwavelength SEDs of the remaining sources in our broad-line AGN sample in Figure \ref{fig:sed_all}, in addition to Figures \ref{fig:quasar_like_SED} and \ref{fig:lrd_like_SED_MIRI}.

\section{Variability of all the samples}\label{sec:variability_appendix}

To optimize image reduction for variability measurements, we reprocess the COSMOS-Web and the COSMOS-3D images following a procedure similar to that described in \S\ref{sec:image_reduction}. For each object, we select all COSMOS-Web and COSMOS-3D F115W exposures (\texttt{\_cal.fits}) that cover the object. We then apply background subtraction and WCS alignment using the existing reference catalog. Afterward, we run \texttt{calwebb\_image3} on the selected exposures and resample them onto the same WCS grid. To further improve alignment, we refine the WCS solution within a 1$\times$1 arcmin$^2$ region around the objects and reproject the images into the same WCS frame again. All configuration parameters for COSMOS-Web and COSMOS-3D image reduction are kept identical to ensure consistency between the two image products. As a result, we obtain two epochs of coadded images for each object. 

We examine the F115W variability through two approaches. (1) We generate differential images by subtracting the COSMOS-Web images from the COSMOS-3D images. In the differential images, we measure the flux using a $r = 0.15\arcsec$ aperture centered on the source position, placing 50 random apertures around it to estimate the uncertainties. The measured flux represents the flux variation between the two epochs.  (2) We directly measure the flux in the COSMOS-Web and COSMOS-3D F115W images using $r = 0.15\arcsec$ apertures and compare the magnitude variation. We list the flux and magnitude variation in Table \ref{tab:var_f115w}.  The details of the differential imaging process are provided in Appendix \ref{sec:variability_appendix}.   For reference, we examine the flux variation of galaxies near our sample. We confirm that the variability of these nearby sources, as determined by both approaches, is consistent with no variation within $1\sigma$.

\begin{table*}[ht!]
	\begin{center}
		\begin{tabular}{cccccc}
        \hline
			ID & $m_{\rm F115W}(r=0.15\arcsec)$ & $\Delta t_{\rm obs}$ & $\Delta t_{\rm rest}$ & $\Delta f_{\rm F115W}(r=0.15\arcsec)$ & $\Delta m_{\rm F115W}(r=0.15\arcsec)$ \\
			
			 & (mag) & (days) & (days) & (nJy) & (mag) \\
            \hline
			27974 & $25.52\pm0.03$ & 351 & 58 & $16.1\pm9.4$ & $-0.11\pm0.06$ \\
			3878 & $28.18\pm0.30$ & 350 & 58 & $-9.9\pm15.0$ & $0.45\pm0.55$ \\
			24359 & $28.21\pm0.28$ & 351 & 58 & $-8.1\pm11.3$ & $0.39\pm0.46$ \\
			20938 & $>28.26$ & 351 & 57 & $-7.6\pm8.6$ & $0.49\pm0.50$ \\
			12739 &  $27.73\pm0.21$ & 356 & 58 & $-6.0\pm12.5$ & $-0.63\pm0.62$ \\
			25997 & $27.44\pm0.19$ & 351 & 57 & $11.0\pm13.8$ & $-0.37\pm0.47$ \\
			16912 & $>27.81$  & 356 & 57 & $-5.5\pm11.3$ & $0.25\pm0.47$ \\
			23915 & $27.12\pm0.16$ & 357 & 56 & $-11.6\pm12.7$ & $0.22\pm0.23$ \\
			17455 & $>27.54$ & 351 & 55 & $-10.0\pm12.8$ & $0.70\pm0.88$ \\
			23756 & $>28.36$ & 351 & 52 & $4.4\pm14.4$ & $-0.36\pm0.90$ \\
			19661 & $27.02\pm0.19$ & 351 & 52 & $4.5\pm11.7$ & $-0.09\pm0.24$ \\
			18221 & $26.44\pm0.18$ & 356 & 50 & $-4.1\pm17.8$ & $0.05\pm0.21$ \\
			13852 & $27.63\pm0.22$ & 715 & 82 & $-0.5\pm9.5$ & $0.02\pm0.31$ \\
        \hline
		\end{tabular}
        \caption{Time interval, flux difference, and magnitude difference between the two epochs of F115W images. $m_{\rm F115W}$ is measured from the COSMOS-3D images using $r = 0.15\arcsec$ apertures. We report the $3\sigma$ upper limits on $m_{\rm F115W}$ if the S/N$<3$. The flux difference, $\Delta f_{\rm F115W}$, is measured from the differential images (COSMOS-3D $-$ COSMOS-Web) using $r = 0.15\arcsec$ apertures. The magnitude difference, $\Delta m_{\rm F115W}$, is derived by comparing the fluxes measured with $r = 0.15\arcsec$ aperture photometry in the two epochs of images. }
        \label{tab:var_f115w}
	\end{center}
\end{table*}

\section{The \ha\ LF}\label{appendix:lf}

\subsection{The maximum survey volume}\label{appendix:vmax}

We first compute the effective survey area for each broad-line AGN, considering regions where the noise levels allow for a reliable identification of the broad \ha\ wings. We generate mock 2D grism spectra for each AGN based on their best-fit models, add noise to the 2D mock spectra, and extract the corresponding 1D spectra. We then fit the extracted 1D spectra using both a single Gaussian model and a two-component Gaussian model. A broad H$\alpha$ line is considered successfully identified if a Gaussian component with FWHM $> 1000$ \si{km\,s^{-1}} is detected with S/N $> 5$, and the model including the broad component yields a reduced $\chi^2$ value at least 0.1 lower than that of the single narrow (FWHM $< 1000$ \si{km\,s^{-1}}) Gaussian model.  We increase the noise levels added to the mock 2D spectra until the broad H$\alpha$ components can no longer be reliably characterized within 500 trials.  We construct the RMS maps using continuum-removed WFSS \texttt{cal} files at redshifts $z = 4.8 - 6.8$ following \cite{Sun2023}, where their H$\alpha$ lines fall within the $3.8 - 5.1$\,\micron\ range, the wavelength range of F444W grism. For each AGN, the maximum sky area at this redshift is determined as the region on the RMS map where the RMS values are smaller than the maximum RMS required for broad-line identification. The maximum survey volume ($V_{\rm max}$) for each object is then integrated across the maximum sky area over the redshift range $z = 4.8 - 6.8$.

\subsection{Completeness correction}

To measure the completeness, we define a series of RMS grids ranging from 0.3 to 50 $\mu$Jy. For each object, we repeat the mock grism experiment for 500 trials as described above: starting with their best-fit model 2D grism spectra, inserting noise at the specific RMS values, and re-performing the extraction and broad-line identification. The completeness at a given RMS value is defined as the fraction of successful broad-line identifications out of the total number of trials. We then count the RMS distribution within the $V_{\rm max}$ for each object, and $C_i$ represents the overall completeness, averaged over the RMS values and their corresponding completeness.  

\subsection{Lynden-Bell's $C^-$ method}\label{sec:appendix_lf_cm}

We estimate the \ha\ LFs using Lynden-Bell $C^-$ method \citep{Lynden-Bell1971, Woodroofe1985, Wang1986}. 
The $C^-$ method is a nonparametric maximum likelihood estimator that accounts for truncation effects, i.e., the boundaries imposed by the selection function on the luminosity--redshift plane in a flux-limited survey. 
It has been previously applied to quasar LFs \citep[e.g.,][]{Fan2001} and is statistically robust against binning effects.

\begin{figure}[h!]
    \centering
    \includegraphics[width=\linewidth]{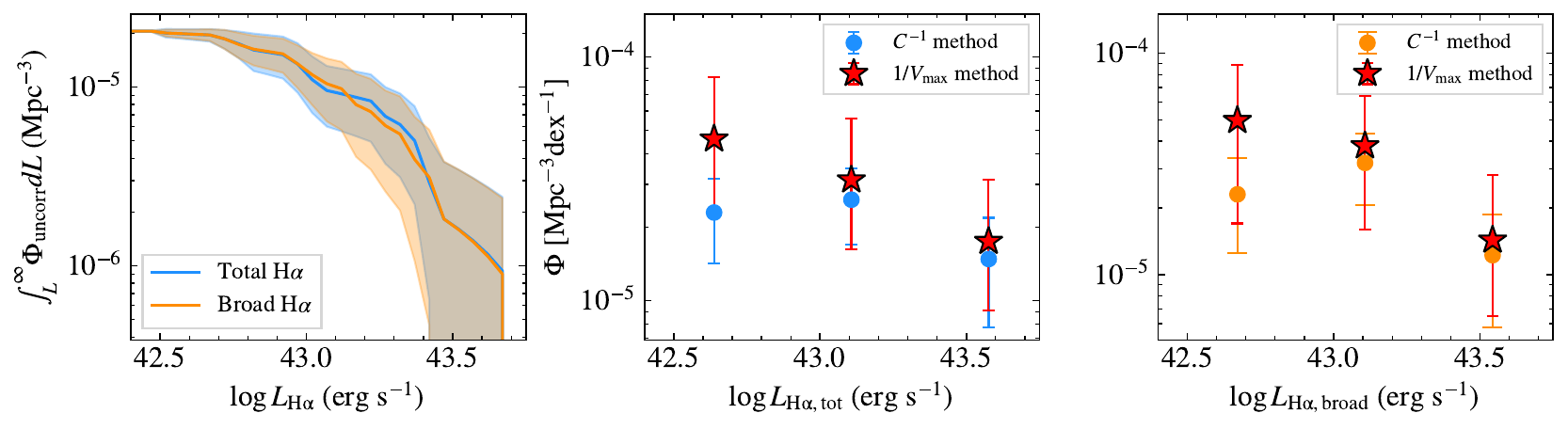}
    \caption{\textit{Left}: Cumulative LF measured using Lynden-Bell's $C^{-1}$ method, uncorrected for completeness. \textit{Middle}: The comparison between the $C^{-1}$ method-measured total H$\alpha$ LF  and 1/$V_{\rm vmax}$ method-measured total H$\alpha$ LF. \textit{Right}:  Comparison between the $C^{-1}$ method-measured broad H$\alpha$ LF  and 1/$V_{\rm vmax}$ method-measured broad H$\alpha$ LF.  }
    \label{fig:cminus}
\end{figure}

We measure the LFs using the $C^-$ estimator\footnote{\url{https://www.astroml.org/modules/generated/astroML.lumfunc.bootstrap_Cminus.html}}. The left panel of Figure \ref{fig:cminus} shows the cumulative LFs uncorrected for completeness. We differentiate the cumulative LFs, then bin the results, apply completeness corrections to each luminosity bin, and estimate the uncertainties through bootstrapping. In Figure \ref{fig:cminus} (middle and right panels) and Table \ref{tab:LF_cm}, we present the differential LFs with three bins. The total \ha\ LF and the broad-line \ha\ LF are both in good agreement within $1\sigma$ with the $1/V_{\rm max}$ results listed in Table \ref{tab:LF}. However, we note the large uncertainties due to the limited sample size. A more statistically robust analysis will be presented in future papers when the full survey data become available.

\begin{table}[h!]
	\begin{center}
		\begin{tabular}{cccc}
        \hline
        \hline
			$\log L_{\rm H\alpha}$ & $\Delta \log L_{\rm H\alpha}$ & N & $\Phi (10^{-5} {\rm Mpc}^{-3} {\rm dex}^{-1})$ \\
			\hline
			$42.638$ & $0.469$ & 4 & $2.00^{+1.02}_{-1.02}$ \\
			$43.107$ & $0.469$ & 4 & $2.95^{+0.98}_{-0.98}$ \\
			$43.576$ & $0.469$ & 4 & $1.45^{+0.72}_{-0.72}$ \\
        \hline
        \hline
        $\log L_{\rm H\alpha, broad}$ & $\Delta \log L_{\rm H\alpha, broad}$ & N & $\Phi (10^{-5} {\rm Mpc}^{-3} {\rm dex}^{-1})$ \\
		\hline
	$42.671$ & $0.436$ & 4 & $2.59^{+1.13}_{-1.13}$ \\
		$43.107$ & $0.436$ & 5 & $3.04^{+1.07}_{-1.07}$ \\
		$43.543$ & $0.436$ & 3 & $1.13^{+0.65}_{-0.65}$ \\
        \hline
		\end{tabular}
    \caption{The \ha\ (narrow + broad) LF and broad-line \ha\ LF measued using Lynden-Bell's $C^-$ method.}
    \label{tab:LF_cm}
	\end{center}
\end{table}

\section{The \hb\ LF}\label{appendix:hb_lf}

\subsection{The \hb\ LF of broad-line AGNs}

The \hb\ LF of broad-line AGNs in Figure \ref{fig:LF_Hb} is measured by combining the $z=7.6$ broad-line AGN (\hbagn) in this work, the two $z=8.5$ broad-line AGNs in \citet{Kokorev2023} and \citet{Fudamoto2025}. 
For the $z=7.6$ source in this work, we follow the procedure outlined in Appendix \ref{appendix:lf}. The maximum volume, $V_{\rm max}$, for this source is $305,418$ Mpc$^{3}$, with an estimated completeness of 0.74. 
The two $z=8.5$ AGNs have comparable intrinsic \hb\ luminosities (lensing magnification corrected for the \citealt{Kokorev2023} source). 
For the \citet{Kokorev2023} source, we assume a maximum survey area of 28\,arcmin$^2$ (UNCOVER Cycle-1 NIRCam primary footprint; \citealt{Bezanson24}) and redshift interval $z=7-9$.
Because of the complexity of NIRSpec target selection and lensing magnification in the Abell 2744 cluster field, the resultant volume is strictly an upper limit. 
For the \citet{Fudamoto2025} source, which was also discovered through NIRCam grism survey, we compute the effective volume using the same method as in \S\ref{sec:ha_lf} and Appendix \ref{appendix:lf}. 
The completeness of these two literature sources is assumed as unity and thus the resultant number density is a lower limit.
We therefore obtain a number density of $\log[\Phi/({\rm Mpc^{-3}\,dex^{-1}})] = -5.26^{+0.76}_{-0.52}$ in the luminosity bin $\log\left[L_{\rm H\beta, tot}/(\,{\rm erg\,s^{-1}})\right] = 43.2 \pm 0.4$ (contributed by ID13852), and $\log[\Phi/({\rm Mpc^{-3}\,dex^{-1}})] >-4.41^{+0.45}_{-0.37}$ in the luminosity bin $\log\left[L_{\rm H\beta, tot}/\,{(\rm erg\,s^{-1})}\right] = 42.4 \pm 0.4$ (contributed by the two AGNs at $z=8.5$ in the literature).

For comparison, we also use the bolometric LF of quasars at $z=8$ as predicted from the TRINITY empirical model \citep{ZhangH2024}.
TRINITY matches the observed quasar LF at $z\lesssim4$ over a wide luminosity range and predicts the LFs at higher redshifts through self consistent empirical modeling of the halo--galaxy--SMBH connection.
For conversion from bolometric to \hb\ luminosity, we use the $L_\mathrm{H\beta} - L_{5100}$ relation measured by \citet{Greene2005} and bolometric correction at rest-frame 5100\,\AA\ measured by \citet{Richards2006}.
These assumptions are similar to those applied to $z\sim5$ quasar \ha\ LF (\S\ref{sec:ha_lf}).

\subsection{The \hb\ LF of star-forming galaxies}

In Figure \ref{fig:LF_Hb}, we also present the \hb\ LF of star-forming galaxies at $z=7-9$ as reference. The JWST Cycle 3 SAPPHIRES EDR \citep{Sun2025} identified seven \hb\ emitters at $z=7-9$ with \hb\ S/N$>4$ over a survey area up to $\sim14$\,arcmin$^2$. We derive the \hb\ LF with this sample following the procedure in Appendix \ref{appendix:lf}, assuming completeness of unity because these source were selected by much brighter [\ion{O}{3}] line emission. 
Given the small survey volume of SAPPHIRES EDR, we also consider the uncertainty propagated from cosmic variance (0.15\,dex) following the \citet{Moster2011} prescription.
For further comparison, we convert the [\ion{O}{3}] LF at $z=7.5-9$ in \cite{Meyer2024} into an \hb\ LF. This conversion adopts the observed \hb/[\ion{O}{3}] ratios as a function of [\ion{O}{3}] luminosity, measured from the JADES DR3 NIRSpec catalog \citep{D'Eugenio2025}. We generate a mock sample of [\ion{O}{3}] emitters and assign \hb\ luminosities through Monte Carlo sampling based on the observed line ratio distribution. This Monte Carl approach provides a converted \hb\ LF that reflects both the underlying [\ion{O}{3}] LF and the empirical \hb--[\ion{O}{3}] flux relation and scattering.
\bigskip

\bibliography{main}{}
\bibliographystyle{aasjournal}

\suppressAffiliationsfalse
\allauthors 
\end{CJK*}
\end{document}